\documentclass[prd,aps,nofootinbib,preprintnumbers,showpacs,showkeys]
{revtex4}
\usepackage{graphicx,epsf,amsfonts,amssymb,amsbsy}
\newcommand{\ds}{\displaystyle}
\newcommand{\vev}[1]{\langle#1\rangle}
\newcommand{\mat}{\left ( \begin{array}}
\newcommand{\emat}{\end{array} \right )}
\newcommand{\vect}{\left ( \begin{array}{c}}
\newcommand{\evect}{\end{array} \right )}

\preprint{HU-EP-10/21}
\begin{document}

\title{ \bf Cooper pairing and finite-size effects in
a NJL-type four-fermion model}
\author{D.~Ebert $^1$ and K. G.~Klimenko $^2$}
\affiliation{$^1$ Institute of Physics, Humboldt-University Berlin,
12489 Berlin, Germany\\
$^2$ IHEP and University "Dubna" (Protvino branch), 142281 Protvino,
Moscow Region, Russia}

\begin{abstract}
Starting from a NJL-type model with N fermion species fermion and
difermion condensates and their associated phase structures are
considered at nonzero chemical potential $\mu$ and zero temperature
in spaces with nontrivial topology of the form $S^1\otimes
S^1\otimes S^1$ and $R^2\otimes S^1$. Special attention is devoted
to the generation of the superconducting phase. In particular, for
the cases of antiperiodic and periodic boundary conditions we
have found that the critical curve of the phase transitions between
the chiral symmetry breaking and superconducting phases as well as
the corresponding condensates and particle densities strongly
oscillate vs $\lambda\sim 1/L$, where $L$ is the length of the
circumference $S^1$. Moreover, it is shown that at some finite
values of $L$ the superconducting phase transition is shifted to
smaller values both of $\mu$ and particle density in comparison with
the case of $L=\infty$.
\end{abstract}


\maketitle
\maketitle

\section{Introduction}

Last years, great theoretical efforts are devoted to the
understanding of the QCD phase diagram. Since at rather small values
of baryonic density weak coupling perturbative QCD methods are not
applicable, usually effective field theories such as the Nambu --
Jona-Lasinio type models (NJL) \cite{njl} etc are invoked for such
kind of investigations
\cite{ebert2,hatsuda,asakawa,hiller,boer,rev}. Evidently, at low
temperatures and baryonic densities one deals with the hadronic
phase. But at growing baryonic density, due to a condensation of
Cooper pairs of two quarks (diquarks), there appears a phase
transition to the so-called color superconducting (CSC) phase of QCD
(see, e.g., the reviews \cite{rev}), where the color symmetry is
spontaneously broken down. In particular, it turned out that
NJL-type models are well-suited for the description of the chiral
symmetry restoring phase transition and low-energy phenomenology of
mesons \cite{2} as well as of properties of CSC quark matter
\cite{rev}.

Since the most attractive feature of NJL models is the dynamical
breaking of the chiral symmetry in the hadronic phase, the
additional influence of different external factors on the chiral
properties of these models was also studied extensively. For
example, they were used to investigate dense baryonic matter in the
presence of external (chromo)magnetic fields \cite{ruggieri}. In
particular, it was demonstrated on the basis of NJL models in
diverse dimensions that both external magnetic \cite{klim} and
chromomagnetic \cite{zhuk} fields induce the chiral symmetry
breaking. Moreover, chiral symmetry breaking in four-fermion models
was studied in weakly curved spaces \cite{odin,eliz} and in spaces
with nontrivial topology, when one or more space coordinates were
compactified \cite{kim,gorbar,gusynin}. 
In addition, the properties of finite size normal quark matter
droplets in the language of the MIT-bag model were investigated,
e.g., in the review \cite{madsen}. Recently, it was also noted that 
the position of the chiral critical end point of the QCD phase
diagram, which could be investigated in heavy ion collision
experiments, depends essentially on the finite system sizes
\cite{fraga}.

There is also some progress in the understanding of the influence of
different external factors on the CSC phase transition. In this
context it is worth mentioning that an external chromomagnetic field
induces the CSC phase transition \cite{ebert} and that an external
magnetic field leads to the appearance of new magnetic phases in 
the three-flavor color superconducting quark matter \cite{incera}.
Moreover, the effect of spaces with constant curvature on CSC was
studied in \cite{etz}. Note also that the stability of finite size
quark matter droplets in the color-flavor locked phase was
investigated in the framework of a bag model using the so-called
multiple expansion method \cite{kiriyama}.

In the present work we shall use an alternative approach in order to
investigate superconductivity in dense cold fermionic matter placed
in a finite volume. In particular, we shall study the Cooper pairing
phenomenon in the framework of a NJL model describing the
interaction of $N$ fermion species in compactified spaces with
nontrivial topology. As in QCD, this model ensures in the usual
$R^3$-space the chiral symmetry breaking at rather small values of
the chemical potential $\mu$, whereas at large values of $\mu$ there
appears a superconducting phase due to the condensation of
difermions.

In this context, let us recall the well-known fact that spontaneous
symmetry breaking in low dimensional quantum field theories may
become impossible due to strong quantum fluctuations of fields
\cite{coleman}. The same is also true for
systems that occupy a limited space volume. However, as it is clear
from physical considerations, the finite size in itself may in some
situations not forbid the spontaneous symmetry breaking, if the
characteristic length of the region of space occupied by the system
is much greater than the Compton wavelength of the excitations
responsible for tunneling and restoration of symmetry. (Indeed, one
may recall here well known physical phenomena such as the
superfluidity of Helium or superconductivity of metals that are
observed in samples of finite volume). This idea has been discussed
for some scalar field theories as well as for NJL models in a closed
Einstein universe, for instance, in \cite{rubakov}. Similarly, if
quantum fluctuations of fields are suppressed when the number of
quantum fields $N$ tends to infinity, spontaneous symmetry breaking
might even occur in a finite volume. Indeed, suppose that in a finite
volume an effective potential of an $O(N)$-symmetrical model has
degenerate global minima. Then, in $D$-dimensional
spacetime the transition probability from one minimum to another is
proportional to $\exp (-NL^{D-2})$ at zero temperature, where $L$ is
the linear size of the system \cite{barducci}. It follows from this
expression that if $L$ and $N$ are finite, the transition
probability is nonzero. This circumstance ensures the vanishing of
the order parameter and, as a result, might lead to a prohibition
for spontaneous symmetry breaking in a finite volume. However, if
$N\to\infty$ the transition probability vanishes and the spontaneous
symmetry breaking is allowed.

In the present paper, the consideration of Cooper pair (difermion)
condensation in restricted regions of space is performed in a toy
NJL model in the mean field approximation, i.e. in the leading order
of the large $N$-expansion technique at $T=0$. In particular, in
order to be sure that the obtained results are stable against
quantum fluctuations, the composite difermion field must be a flavor
singlet (like the composite fermion-antifermion field) which
technically leads to an $N$-factor in the part of the effective
action arising from fermion loops and thus guarantees the
application of the $1/N$ expansion. It is just by this reason that
we demand here $O(N)$ flavor symmetry as opposed to the usual
$SU(N)$ symmetry. \footnote{Note the important difference to the
case of QCD-like NJL models with $SU(N)$ color symmetry, where the
usual $1/N$ expansion cannot be applied to {\it colored diquarks} due
to the lack of a corresponding $N$ factor from quark loops.}

The paper is organized as follows. For comparision, we first derive
in Section II the expression for the thermodynamic potential of cold
dense fermionic matter, described by a NJL-type Lagrangian, for the
case of $R^3$ space. It is shown here that for some fixed values of
coupling constants two phases are allowed, the chiral symmetry
breaking phase (at $\mu<\mu_c\approx 0.3$ GeV) and the
superconducting one (at $\mu>\mu_c$). In Section III the phase
structure of the model is investigated in the  $S^1\otimes
S^1\otimes S^1$ space with periodic and antiperiodic boundary
conditions for fermion fields. We have found a rather rich phase
structure in the $(\mu,\lambda\sim 1/L)$-plane, where $L$ is the
lenght of each circumference $S^1$. It turns out that the boundary
between the chiral symmetry breaking and superconducting phases as
well as the corresponding condensate values and particle densities
strongly oscillate vs $\lambda$ at $\lambda\to 0$. Finally, in
Section IV similar considerations were performed in the case of
$R^2\otimes S^1$ space topology, where we have found smoother
oscillations of both the critical curve and the condensates vs
$\lambda$.

\section{The case of $R^3$ space}

\subsection{The model and its thermodynamic potential}

Our investigation is based on an NJL--type model with massless
fermions belonging to a fundamental multiplet of the $O(N)$ flavor
group. Its Lagrangian describes the interaction in the
fermion--antifermion as well as scalar difermion channels:
\begin{eqnarray}
 L=\sum_{k=1}^{N}\bar \psi_k\Big [\gamma^\nu i\partial_\nu
+\mu\gamma^0\Big ]\psi_k+ \frac GN\left (\sum_{k=1}^{N}\bar
\psi_k\psi_k\right )^2+\frac HN\left (\sum_{k=1}^{N}\bar
\psi_k^Ci\gamma^5\psi_k\right )\left (\sum_{j=1}^{N}\bar
\psi_ji\gamma^5\psi_j^C\right ), \label{1}
\end{eqnarray}
where $\mu$ is a fermion number chemical potential. As it is noted
above, all fermion fields $\psi_k$ ($k=1,...,N$) form a fundamental
multiplet of $O(N)$ group. Moreover, each field $\psi_k$ is a
four-component Dirac spinor; $\psi_k^C=C\bar \psi_k^t$ and $\bar
\psi_k^C=\psi_k^t C$ are charge-conjugated spinors, and
$C=i\gamma^2\gamma^0$ is the charge conjugation matrix (the symbol
$t$ denotes the transposition operation).  Clearly, the Lagrangian
$L$ is invariant under transformations from the internal $O(N)$
group, which is introduced here in order to make it possible to
perform all the calculations in the framework of the nonperturbative
large-$N$ expansion method. Physically more interesting is that the
model (1) is invariant under transformations from an Abelian
electric charge $U(1)$ group: $\psi_k\to\exp{i\alpha}\psi_k$
($k=1,...,N$). In addition, the Lagrangian is invariant under the
discrete $\gamma^5$ chiral transformation:
$\psi_k\to\gamma^5\psi_k$, $\bar\psi_k\to-\bar\psi_k\gamma^5$
($k=1,...,N$). The linearized version of Lagrangian (\ref{1}) that
contains auxiliary scalar bosonic fields $\sigma (x)$, $\Delta(x)$,
$\Delta^{*}(x)$ has the following form
\begin{eqnarray}
{\cal L}\ds =\bar\psi_k\Big [\gamma^\nu i\partial_\nu
+\mu\gamma^0 -\sigma \Big ]\psi_k
 -\frac{N}{4G}\sigma^2 -\frac N{4H}\Delta^{*}\Delta-
 \frac{\Delta^{*}}{2}[\bar\psi_k^Ci\gamma^5\psi_k]
-\frac{\Delta}{2}[\bar\psi_k i\gamma^5\psi_k^C].
\label{3}
\end{eqnarray}
(Here and in the following summation over repeated indices
$k=1,...,N$ is implied.) Clearly, the Lagrangians (\ref{1}) and
(\ref{3}) are equivalent, as can be seen by using the Euler-Lagrange
equations of motion for scalar bosonic fields $\sigma (x)$,
$\Delta(x)$, $\Delta^{*}(x)$, which take the form 
\begin{eqnarray}
\sigma (x)=-2\frac GN(\bar\psi_k\psi_k),~~
\Delta(x)=-2\frac HN(\bar
\psi_k^Ci\gamma^5\psi_k),~~
\Delta^{*}(x)=-2\frac HN(\bar\psi_k i\gamma^5\psi_k^C).
\label{4}
\end{eqnarray}
One can easily see from (\ref{4}) that the (neutral) field
$\sigma(x)$ is a real quantity, i.e. $(\sigma(x))^\dagger=\sigma(x)$
(the superscript symbol $\dagger$ denotes the hermitian
conjugation), but the (charged) difermion field $\Delta(x)$ is a
complex scalar, so $(\Delta(x))^\dagger= \Delta^{*}(x)$. 
Clearly, all the fields (\ref{4}) are singlets with respect to the
$O(N)$ group. \footnote{Note
that the $\Delta (x)$ field is a flavor O(N) singlet, since the
representations of this group are real.} If the scalar difermion
field $\Delta(x)$ has a nonzero
ground state expectation value, i.e.\  $\vev{\Delta(x)}\ne 0$, the
Abelian $U(1)$ charge symmetry of the model is spontaneously broken
down. However, if $\vev{\sigma (x)}\ne 0$ then the discrete chiral
symmetry of the model is spontaneously broken.

Let us now study the phase structure of the four-fermion model (1)
by starting with the equivalent semi-bosonized Lagrangian (\ref{3}).
In the leading order of the large-$N$ approximation, the effective
action ${\cal S}_{\rm {eff}}(\sigma,\Delta,\Delta^{*})$ of the
considered model is expressed by means of the path integral over
fermion fields:
$$
\exp(i {\cal S}_{\rm {eff}}(\sigma,\Delta,\Delta^{*}))=
  \int\prod_{l=1}^{N}[d\bar\psi_l][d\psi_l]\exp\Bigl(i\int {\cal
  L}\,d^4 x\Bigr),
$$
where
\begin{eqnarray}
&&{\cal S}_{\rm {eff}}
(\sigma,\Delta,\Delta^{*})
=-\int d^4x\left [\frac{N}{4G}\sigma^2(x)+
\frac{N}{4H}\Delta (x)\Delta^{*}(x)\right ]+
\widetilde {\cal S}_{\rm {eff}}.
\label{5}
\end{eqnarray}
The fermion contribution to the effective action, i.e.\  the term
$\widetilde {\cal S}_{\rm {eff}}$ in (\ref{5}), is given by:
\begin{equation}
\exp(i\widetilde {\cal S}_{\rm
{eff}})=\int\prod_{l=1}^{N}[d\bar\psi_l][d\psi_l]\exp\Bigl(\frac{i}{2
}\int\Big [\bar\psi_k D^+\psi_k+\bar\psi_k^CD^-\psi_k^C-\bar\psi_k
K\psi_k^C-\bar \psi_k^CK^{*}\psi_k\Big ]d^4 x\Bigr), \label{6}
\end{equation}
where we have used the following notations \footnote{In order to
bring the fermion sector of the Lagrangian (\ref{3}) to the
expression, given in the square brackets of (\ref{6}), we use the
following well-known relations: $\partial_\nu^t=-\partial_\nu$,
$C\gamma^\nu C^{-1}=-(\gamma^\nu)^t$,
$C\gamma^5C^{-1}=(\gamma^5)^t=\gamma^5$.}
\begin{eqnarray}
&&D^\pm=i\gamma^\nu\partial_\nu\pm\mu\gamma^0-\sigma (x),~~~
K^*=i\Delta^{*}(x)\gamma^5,\qquad
K=i\Delta(x)\gamma^5.
\label{7}
\end{eqnarray}
In the following, it is very convenient to use the Nambu--Gorkov
formalism, in which for each fixed $k=1,...,N$ a pair of fermion
fields $\psi_k$ and $\psi_k^C$ are composed into a bispinor $\Psi_k$
such that
\begin{equation}
\Psi_k=\left({\psi_k\atop\psi_k^C}\right),~~\Psi_k^t=(\psi_k^t,\bar
\psi_k C^t);~~
\quad \bar\Psi_k=(\bar\psi_k,\bar\psi_k^C)=(\bar\psi_k,\psi_k^t
C)=\Psi_k^t \left
(\begin{array}{cc}
0~~,&  C\\
C~~, &0
\end{array}\right )\equiv\Psi_k^t Y.
\label{8}
\end{equation}
Furthermore, by introducing the matrix-valued operator
\begin{equation}
Z=\left (\begin{array}{cc}
D^+, & -K\\
-K ^*, &D^-
\end{array}\right ),\label{9}
\end{equation}
one can rewrite the gaussian functional integral in (\ref{6}) in
terms of $\Psi_k$ and $Z$ and then evaluate it as follows (clearly,
in this case $[d\bar\psi_k][d\psi_k]=$
$[d\psi_k^C][d\psi_k]=$$[d\Psi_k]$):
$$
\exp(i\tilde {\cal S}_{\rm {eff}})=
\int\prod_{l=1}^{N}[d\Psi_l]\exp\left\{\frac
i2\int\bar\Psi_k Z\Psi_k d^4x\right\}=
  \int\prod_{l=1}^{N}[d\Psi_l]\exp\left\{\frac
  i2\int\Psi_k^t(YZ)\Psi_k
  d^4x\right\}=\mbox {det}^{N/2}(YZ)=\mbox {det}^{N/2}(Z),
$$
where the last equality is valid due to the evident relation $\det
Y=1$. Then, using the relation (\ref{5}), 
one obtains the expression for the effective action:
\begin{equation}
{\cal S}_{\rm {eff}}(\sigma,\Delta,\Delta^{*})
=-\int d^4x\left [\frac{N}{4G}\sigma^2(x)+
\frac{N}{4H}\Delta (x)\Delta^{*}(x)\right ]-i\frac N2\ln\mbox
{det}(Z)
\label{10}
\end{equation}
Starting from (\ref{10}), one can define the thermodynamic potential
(TDP) $\Omega(\sigma,\Delta,\Delta^{*})$ of the model (\ref{1}) in
the leading order of the large-$N$ expansion (mean field
approximation):
\begin{equation}
{\cal S}_{\rm {eff}}~\bigg |_{~\sigma,\Delta,\Delta^{*}=\rm {const}}
=-N\Omega(\sigma,\Delta,\Delta^{*})\int d^4x. \label{11}
\end{equation}
Here we have supposed that the quantities $\sigma,\Delta,\Delta^{*}$
do not depend on coordinates $x$. Moreover, without loss of
generality, one can set the arbitrary phase of $\Delta$ equal to
zero so that $\Delta$ is now considered as a non-negative real
quantity, i.e. $\Delta$=$|\Delta|$. As a consequence, the det$Z$ in
(\ref{10}) and the TDP (\ref{11}) are easily calculated. Indeed,
using the general formulae
$$
\det\left
(\begin{array}{cc}
A~, & B\\
\bar A~, & \bar B
\end{array}\right )=\det [-\bar AB+\bar AA\bar A^{-1}\bar B]
$$
and $\det O=\exp {\rm Tr}\ln O$, one can find for the TDP (\ref{11})
the expression:
\begin{eqnarray}
\Omega(\sigma,\Delta)=\frac{\sigma^2}{4G}+
\frac{\Delta^2}{4H}+\frac i2\frac{{\rm
Tr}_{sx}\ln D}{\int d^4x},
\label{12}
\end{eqnarray}
where $D=\Delta^2+\gamma^5D^+\gamma^5D^-$ and the Tr-operation
stands for the trace in spinor ($s$) and four-dimensional coordinate
($x$) spaces, respectively. Transferring in (\ref{12}) to the
momentum space representation for the operator $D$, we have
\begin{eqnarray}
\Omega(\sigma,\Delta)=\frac{\sigma^2}{4G}+
\frac{\Delta^2}{4H}+\frac i2{\rm
Tr}_{s}\int\frac{d^4p}{(2\pi)^4}\ln\Big [\Delta^2+\gamma^5(\not\!p
+\mu\gamma^0-\sigma)\gamma^5(\not\!p -\mu\gamma^0-\sigma)\Big ],
\label{13}
\end{eqnarray}
where in the square brackets just the momentum space representation,
$\bar D$, for the operator $D$ appears. In  four-dimensional spinor
space the 4$\times$4 matrix $\bar D$ has two different eigenvalues
$\epsilon_\pm$, each being two-fold degenerate:
\begin{eqnarray}
\epsilon_\pm=(E_\Delta^\pm)^2-p_0^2\equiv
(E\pm\mu)^2+\Delta^2-p_0^2, \label{14}
\end{eqnarray}
where $E=\sqrt{\sigma^2+\vec p^2}$. Since ${\rm Tr}_{s}\ln\bar
D=2\ln\epsilon_+\epsilon_-$, one can integrate in (\ref{13}) over
$p_0$ and obtain the following expression for the TDP of dense cold
fermion matter (more details of this technique are presented, e.g.,
in \cite{tyuk}):
\begin{eqnarray}
\Omega(\sigma,\Delta)=\frac{\sigma^2}{4G}+
\frac{\Delta^2}{4H}-\int\frac{d^3p}{(2\pi)^3}\Theta (\Lambda^2 -\vec
p^2 )\Big [E_\Delta^++E_\Delta^-\Big ], \label{15}
\end{eqnarray}
where the Heaviside step-function $\Theta (x)$ has been inserted in
order to regularize the ultraviolet divergent integral, and
$\Lambda$ is a cutoff parameter that is usually taken smaller than 1
GeV, i.e. $\Lambda<1$ GeV. Since the TDP (\ref{15}) is symmetric
with respect to the transformations $\sigma\to -\sigma$, and $\mu\to
-\mu$, we suppose in the following that $\sigma\ge 0$ and $\mu\ge 0$
(recall also $\Delta\ge 0$). It is important to note that the
quantities $E_\Delta^-$ and $E_\Delta^+$ defined in (\ref{14}) are
the energies of
fermions and antifermions (quasiparticles) in a medium,
correspondingly. Clearly, each energy level is infinitely
degenerated with respect to the direction of the momentum $\vec p$.
Indeed, there are infinitely many quasiparticles with the same
energy but with different directions of momenta.

\begin{figure}
 \includegraphics[width=0.45\textwidth]{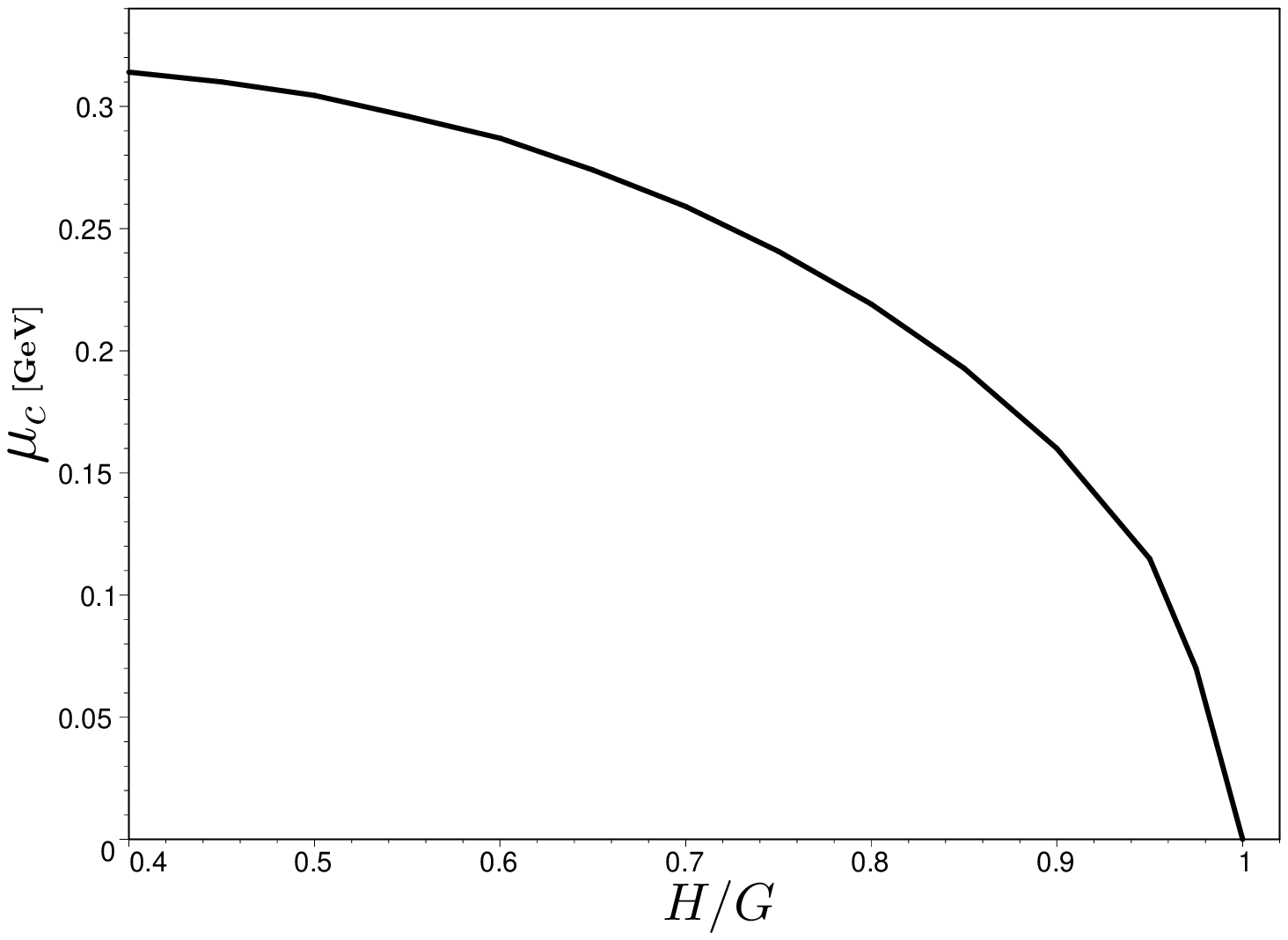}
 \hfill
 \includegraphics[width=0.45\textwidth]{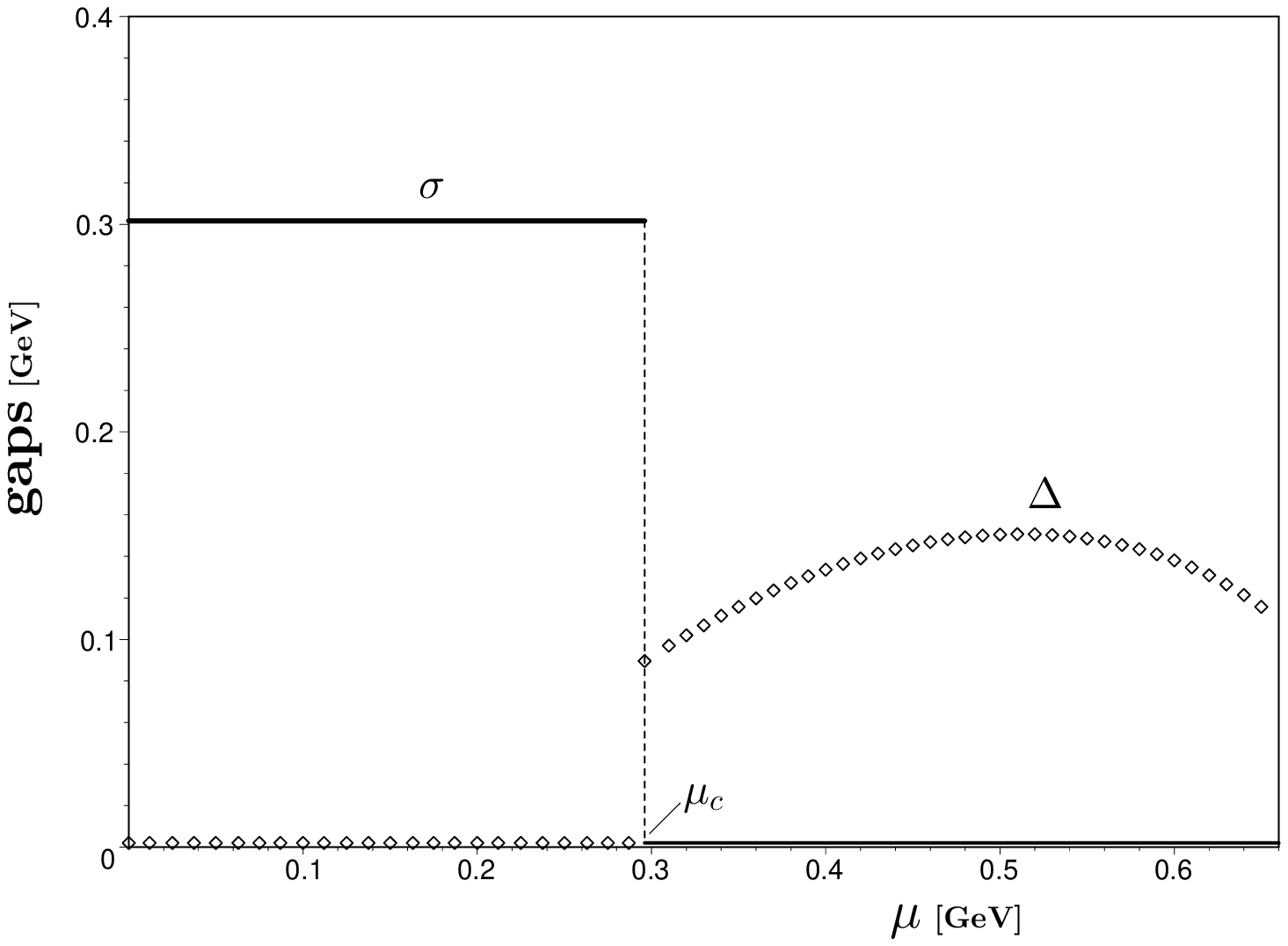}\\
\parbox[t]{0.45\textwidth}{ \caption{The case of $R^3$ space
topology: The behavior of the critical value of the chemical
potential $\mu_c$ vs $H$ at $G=30.06$~GeV$^{-2}$.}
 }\hfill
 \parbox[t]{0.45\textwidth}{
\caption{The case of $R^3$ space topology: The gaps $\sigma$ and
$\Delta$ vs $\mu$, where $\mu_c\approx 296$ MeV.} }
\end{figure}

\subsection{Phase structure}

In order to obtain the phase structure of the initial model it is
necessary to investigate the behavior of the global minimum point
(GMP) of the TDP (\ref{15}) in dependence on the chemical potential
$\mu$. The coordinates of this point are usually called gaps. (The
$\sigma$- and $\Delta$-coordinates of the GMP are the chiral and
difermion condensates, respectively.) In the model two types of the
GMPs are allowed, i) $(\sigma\ne 0,0)$ and ii) $(0,\Delta\ne 0)$.
The GMP of the i)-th type corresponds to the phase with broken
chiral $\gamma^5$-invariance only (it is a so-called normal phase),
whereas the GMP of the type ii) corresponds to the superconducting
phase, in which Cooper pairing of fermions leads to the spontaneous
breaking of the $U(1)$ symmetry.

Throughout the paper, we use in our numerical calculations the value
$G=30.06$ GeV$^{-2}$. Moreover, for illustrations, let us take the
cutoff in the momentum integral in (\ref{15}) to be $\Lambda =650$
MeV. For this choice of parameters, the TDP (\ref{15}) then predicts
at $\mu=0$ and $H=0$ (the last is equivalent to the constraint
$\Delta=0$) a corresponding value of the chiral condensate which is
characteristic to some low energy phenomenological QCD-like NJL
models \cite{rev}. In the case under consideration the phase
structure of the model depends essentially on the value of the
coupling constant $H$. Numerical calculations show that at $H<G$ and
sufficiently low values of $\mu < \mu_c$ the GMP of the TDP
(\ref{15}) corresponds to the normal phase of the model. However, at
$\mu >\mu_c$ the superconducting phase is realized. If $H>G$, then
for all values of $\mu$, and even for $\mu=0$, the ground state of
the system corresponds to the superconducting phase. The behavior of
the critical value of the chemical potential $\mu_c$ vs $H$ is
depicted in Fig. 1. (At the same time, one may consider Fig. 1 as a
phase portrait of the model in terms of $\mu$ and $H$. Then, below
(above) the critical line $\mu_c$ the normal (superconducting) phase
of the model is realized.) In all subsequent numerical calculations
the coupling constant $H$ is fixed by the relation $H=0.55~ G$.
Hence, the set of model parameters in our investigations is the
following:
\begin{eqnarray}
G=30.06~\mbox{GeV}^{-2}, ~~~H=0.55~ G,~~\Lambda =650~\mbox{MeV}.
\label{2}
\end{eqnarray}
As a result, we have for the set (\ref{2}) $\mu_c\approx 0.3$ GeV.
Moreover, the behavior of gaps $\sigma$ and $\Delta$ vs $\mu$ in
this case is presented in Fig. 2. It is clear from this figure that
at $\mu<\mu_c$ the GMP of the TDP (\ref{15}) has the form
$(\sigma,0)$ (as a consequence, the system is in the normal phase),
where $\sigma\approx 0.3$~GeV, whereas for $\mu>\mu_c$ the
superconducting phase, corresponding to the GMP of the form
$(0,\Delta)$ of the TDP, is realized in the system. Evidently, in
the critical point $\mu_c$ there is a first order phase transition.

\section{The case of  $S^1\otimes S^1\otimes S^1$ space topology}

In the present section we generalize the previously obtained results
to the case of a space with finite volume. Evidently, this is a
reasonable task, since all physical effects take place in restricted
space regions. For simplicity, let us suppose that our system is
immersed into a box with equal linear sizes, $0\le x,y,z\le L$. It
is well known that in this case the task is equivalent to the
consideration of the model in the space of nontrivial $S^1\otimes
S^1\otimes S^1$ topology with quantum fields satisfying some
boundary conditions of the form (here we again simplify the problem,
demanding identical boundary conditions for all coordinates):
\begin{eqnarray}
\psi_k(t,x+L,y,z)=e^{i\pi\alpha}\psi_k(t,x,y,z),~~
\psi_k(t,x,y+L,z)=e^{i\pi\alpha}\psi_k(t,x,y,z),~~
\psi_k(t,x,y,z+L)=e^{i\pi\alpha}\psi_k(t,x,y,z), \label{16}
\end{eqnarray}
where $0\le\alpha < 2$, $L$ is the length of the circumference
$S^1$, and now each of the variables $x,y,z$ mean the path along it.
Below, we shall use only two values of the parameter $\alpha$:
$\alpha=0$ for periodic boundary conditions and $\alpha=1$ for the
antiperiodic one.

As a consequence, to obtain the thermodynamic potential
$\Omega_{L\alpha}(\sigma,\Delta)$ of fermions moving in a
space with nontrivial topology $S^1\otimes S^1\otimes S^1$, we
should replace the integration over each momentum in (\ref{15})
by a summation over corresponding discrete momenta
$p_{n\alpha}$ following the rule:
\begin{eqnarray}
\int\frac{d^3p}{(2\pi)^3}f(p_{x},p_y,p_z)\to\frac
1{L^3}\sum_{k=-\infty}^{\infty}\sum_{l=-\infty}^{\infty}
\sum_{m=-\infty}^{\infty}
f(p_{k\alpha},p_{l\alpha},p_{m\alpha}),~~~~p_{n\alpha}=
\frac{\pi}{L}(2n+\alpha),~~~n=0,\pm 1, \pm 2,... \label{17}
\end{eqnarray}

\subsection{The case of antiperiodic boundary conditions}

Applying the rule (\ref{17}) with $\alpha =1$ in the expression
(\ref{15}), one immediately obtains the TDP of the system in the
case of antiperiodic ( ``a'') boundary conditions:
\begin{eqnarray}
\Omega_{L a}(\sigma,\Delta)&=&
\frac{\sigma^2}{4G}+\frac{\Delta^2}{4H}-
\frac{8\lambda^3}{\pi^3}\sum_{i=0}^\infty\sum_{k=0}^\infty\sum_{l=0}^
\infty
\Theta (\Lambda^2-p^2_{ia}-p^2_{ka}-p^2_{la})\Big [{\cal E}_{\Delta
ikl}^{a+}+ {\cal E}_{\Delta ikl}^{a-}\Big ],\label{18}
\end{eqnarray}
where $\lambda=\pi/L$, ${\cal E}_{\Delta ikl}^{a\pm}=\sqrt{\left
({\cal E}_{ikl}\pm\mu\right )^2+\Delta^2}$, ${\cal
E}_{ikl}=\sqrt{p^2_{ia}+p^2_{ka}+p^2_{la}+\sigma^2}$, and
$p_{ia}=\lambda (2i+1)$ etc.
\begin{figure}
 \includegraphics[width=0.45\textwidth]{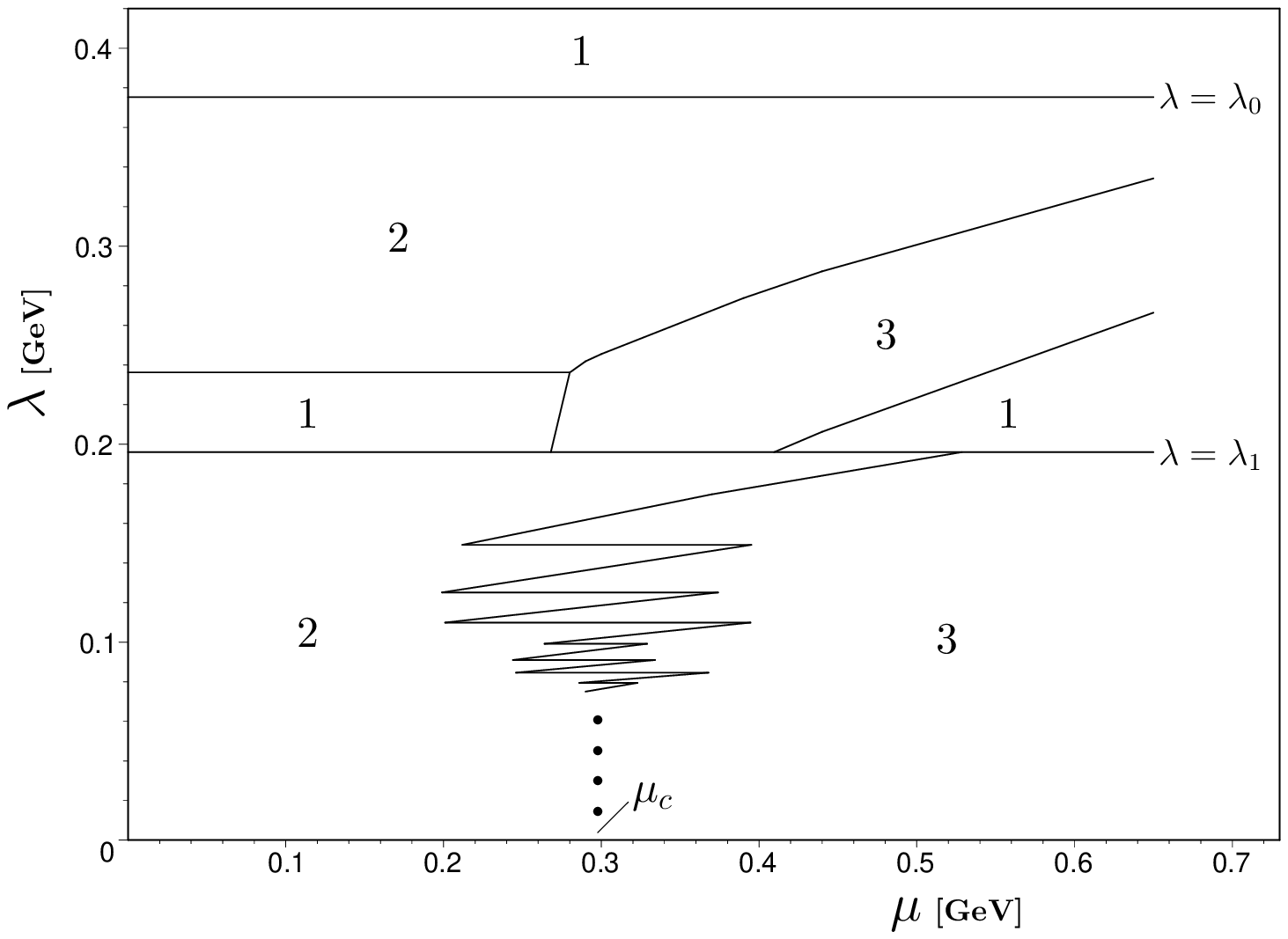}
 \hfill
 \includegraphics[width=0.45\textwidth]{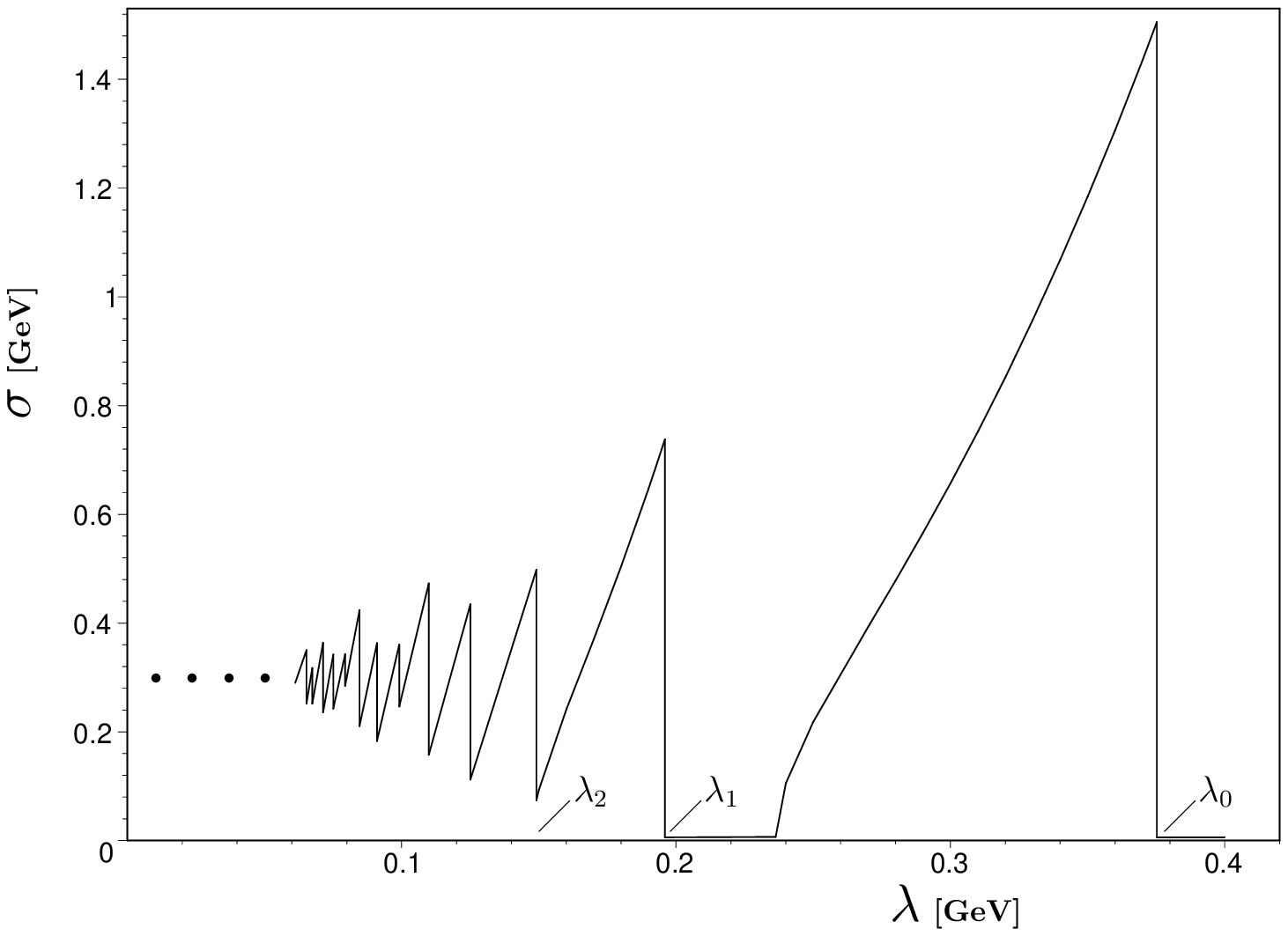}\\
\parbox[t]{0.45\textwidth}{ \caption{Phase structure in the case of
$S^1\otimes S^1\otimes S^1$ space topology with antiperiodic
boundary conditions. The numbers 1, 2 and 3 denote the symmetric
phase with $\sigma=0,\Delta=0$, the chirally broken phase with
$\sigma\ne 0,\Delta=0$ and the superconducting phase with $\sigma=
0,\Delta\ne 0$, correspondingly. Here $\lambda=\pi/L$, $\mu_c\approx
0.296$ GeV, $\lambda_0\approx 0.375$ GeV, $\lambda_1\approx 0.195$
GeV.}
 }\hfill
 \parbox[t]{0.45\textwidth}{\caption{The gap $\sigma$
 vs $\lambda=\pi/L$ at $\mu=0.18$ GeV in the case of  $S^1\otimes
 S^1\otimes S^1$ space topology with antiperiodic boundary
 conditions. Here $\lambda_0\approx 0.375$ GeV, $\lambda_1\approx
 0.195$ GeV, $\lambda_2\approx 0.149$ GeV. } }
\end{figure}

The quantities ${\cal E}_{\Delta ikl}^{a\pm}$ in (\ref{18}) are the
energies of elementary one-fermion excitations (quasiparticles) in a
dense medium which occupies now a finite volume and is constrained
by antiperiodic boundary conditions. (The signs -/+ correspond to
the energies of fermion/antifermion quasiparticles.) Evidently, both
fermion and antifermion quasiparticle energy levels can be labeled
by a triple of integers, $(i,k,l)$, where $i\ge k\ge l\ge 0$.
Clearly, in a finite volume the degeneracy of quasiparticle energy
levels are partially (or even totally) removed. For example, each
energy level with quantum numbers $(0,0,0)$ is non-degenerated, the
level $(1,0,0)$ is three-fold degenerated etc. Now, for each energy
level $(i,k,l)$ let us put into correspondence the integer
$N_{ikl}=(2i+1)^2+(2k+1)^2+(2l+1)^2$ and, as a result, the real
quantity $\lambda_{ikl}\equiv\sqrt{\Lambda^2/N_{ikl}}$. \footnote{It
might occur that for energy levels with different quantum numbers
$(i,k,l)$ there arises the same integer $N_{ikl}$. For example, for
the energy levels with quantum numbers $(1,1,1)$ and $(2,0,0)$ we
have $N_{111}$=$N_{200}=27$ etc. } Using these definitions, we
construct an infinite set of real scales of the form
\begin{eqnarray}
\lambda_0>\lambda_1>\lambda_2>\cdots>\lambda_n>\cdots,
\label{19}
\end{eqnarray}
where each scale $\lambda_k$ coincides with one of the above
obtained expressions $\lambda_{ikl}$ and vice versa, each quantity
$\lambda_{ikl}$ is equal to some element of the sequence (\ref{19}).
So, $\lambda_0=\sqrt{\Lambda^2/3}=\lambda_{000}\approx 0.375$ GeV,
$\lambda_1=\sqrt{\Lambda^2/11}=\lambda_{100}\approx 0.195$ GeV,
$\lambda_2=\sqrt{\Lambda^2/19}=\lambda_{110}\approx 0.149$ GeV,
$\lambda_3=\sqrt{\Lambda^2/27}=\lambda_{111}=\lambda_{200}\approx
0.125$ GeV, $\lambda_4=\sqrt{\Lambda^2/35}$ etc.

Recall that our aim is to investigate the phase structure of the
system with the TDP presented in (\ref{18}). This means that it is
necessary to study the behavior of the global minimum point (GMP) of
the TDP vs $\mu$ and $\lambda$ (or $L$). The structure of the TDP
(\ref{18}) suggests the following strategy for studying the
corresponding GMP. Let us divide the plane $(\mu,\lambda)$
into an infinite set of strips, parallel to the $\mu$-axis:
\begin{eqnarray}
\omega_0=\{(\mu,\lambda):~\lambda>\lambda_0\},~\omega_1=\{(\mu,
\lambda): ~\lambda_0>\lambda>\lambda_1\},~\cdots~,~
\omega_n=\{(\mu,\lambda):~\lambda_{n-1}>\lambda>\lambda_n\},~\dots
\label{20}
\end{eqnarray}
Due to the presence of the theta-function in (\ref{18}), each sum is
indeed a finite one there. Furthermore, for the values of
$(\mu,\lambda)$ from the strip $\omega_0$ the argument of the
theta-function is negative and hence the term with sums vanishes
there. We see that in this case the TDP is reduced to the quantity
$\Omega_{La0}=(\sigma^2/4G+\Delta^2/4H)$ whose minimum lies at the
point $(\sigma =0,\Delta =0)$. As a result, all the points of the
strip $\omega_0$ correspond to the symmetric phase of the model (see
Fig. 3). If the point $(\mu,\lambda)$ belongs to the strip
$\omega_1$, then only the energy levels with quantum numbers
$(0,0,0)$ contribute to the sum in (\ref{18}), so the TDP (\ref{18})
reduces to the quantity $\Omega_{La1}$,
\begin{eqnarray}
\Omega_{La1}=\frac{\sigma^2}{4G}+\frac{\Delta^2}{4H}-
\frac{8\lambda^3}{\pi^3}\left [\sqrt{\left
(\sqrt{\sigma^2+3\lambda^2}+\mu\right )^2+\Delta^2}+\sqrt{\left
(\sqrt{\sigma^2+3\lambda^2}-\mu\right )^2+\Delta^2}\right ].
\label{21}
\end{eqnarray}
The form of the GMP of this function depends essentially on the
values of $(\mu,\lambda)$, so in the strip $\omega_1$, as numerical
calculations show, there are three different phases (see Fig. 3):
the symmetric phase 1 corresponding to the GMP of the form
$(\sigma=0, \Delta=0)$, the chirally broken phase 2 with $(\sigma\ne
0,\Delta=0)$ and superconducting (SC) phase 3 with $(\sigma=
0,\Delta\ne 0)$.
\begin{figure}
 \includegraphics[width=0.45\textwidth]{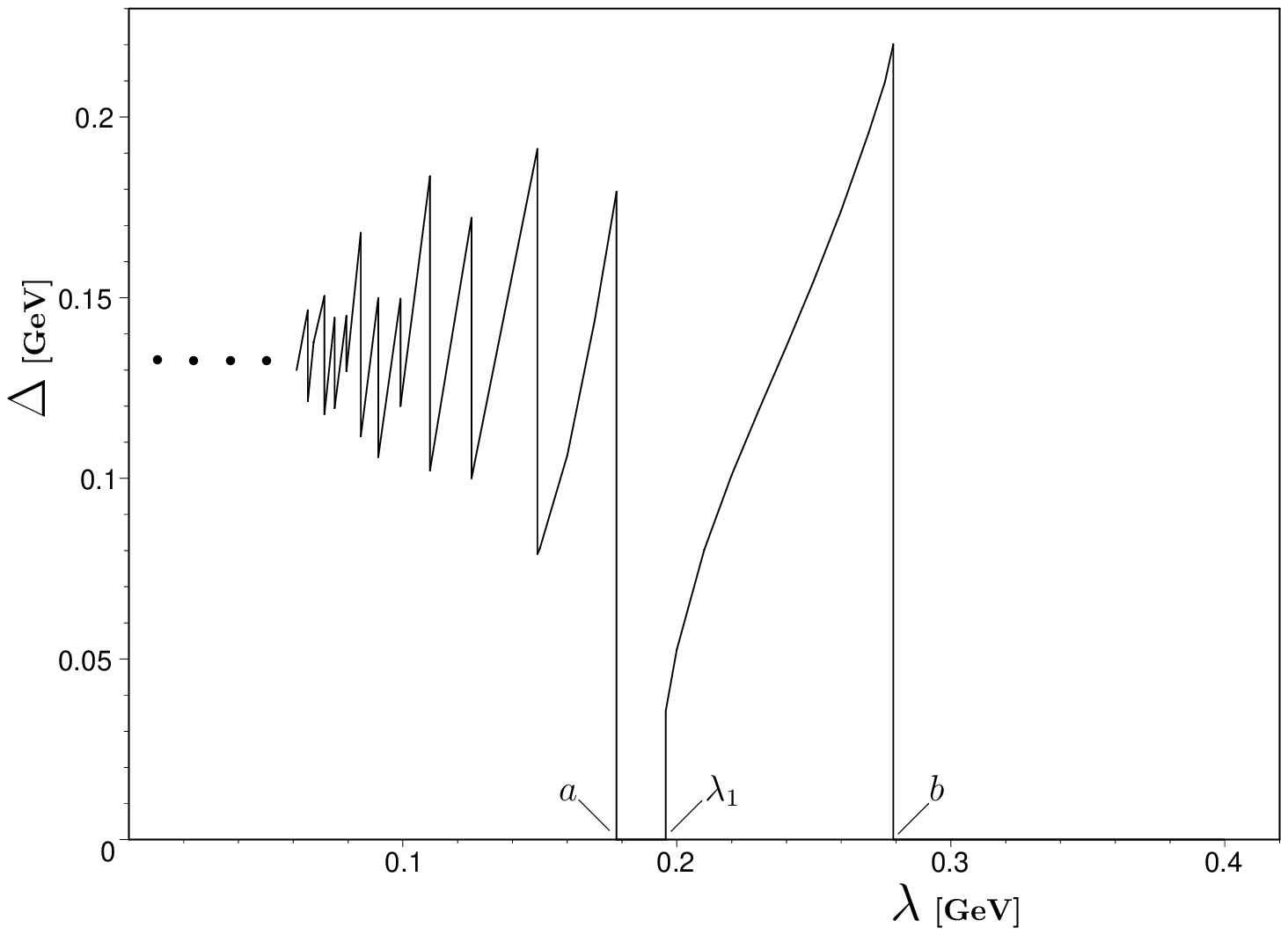}
 \hfill
 \includegraphics[width=0.45\textwidth]{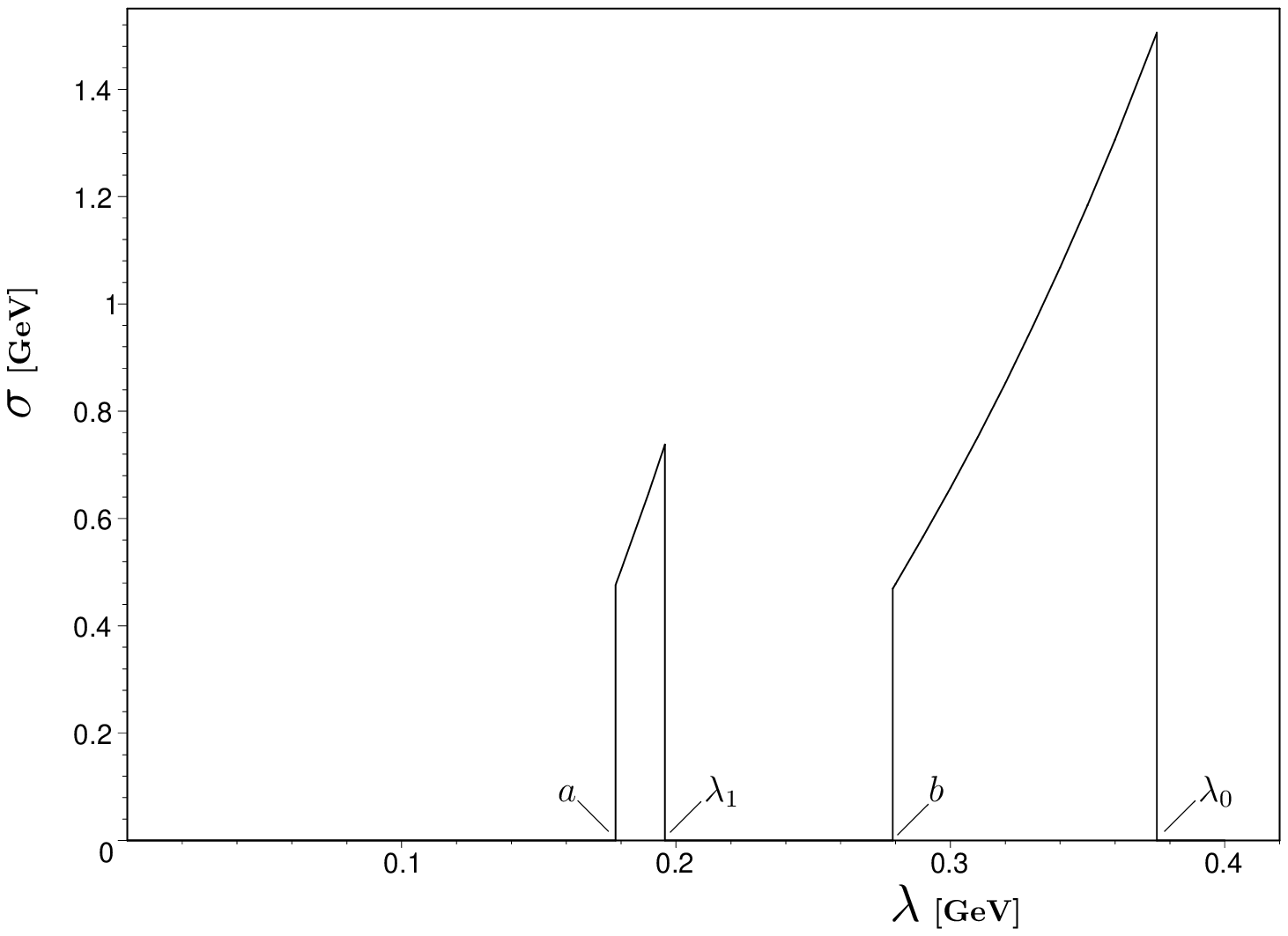}\\
\parbox[t]{0.45\textwidth}{ \caption{The gap $\Delta$
vs $\lambda$ at $\mu=0.4$ GeV  in the case of  $S^1\otimes
S^1\otimes S^1$ space topology with antiperiodic boundary conditions.
Here $a\approx 0.178$ GeV, $b\approx 0.279$ GeV and other notations
are presented in Figs 3, 4. }
 }\hfill
 \parbox[t]{0.45\textwidth}{\caption{The gap $\sigma$
 vs $\lambda$ at $\mu=0.4$ GeV in the case of  $S^1\otimes
 S^1\otimes  S^1$ space topology with antiperiodic boundary
 conditions. The notations are the same as in Figs 3-5.
 } }
\end{figure}

To study the phase structure of the model in the strip $\omega_2$,
it is necessary to take into account in the sums (\ref{18}) the
contribution from the energy levels $(1,0,0)$ in addition. In the
strip $\omega_3$ the energy levels with quantum numbers $(1,1,0)$
should be switched on in addition to the previous ones, and so on.
However, in each of the strips $\omega_k$ with $k\ge 2$ only the
phases 2 and 3 might exist, so for all $\lambda <\lambda_1$ we have
found in the plane $(\mu,\lambda)$ just these two phases, the phase
with broken chiral symmetry and the superconducting one. In Fig. 3
they are arranged below the line $\lambda=\lambda_1$ and divided by
a zigzag, or oscillating, critical line. The amplitude of
oscillations of this line is rather large for values of $\lambda$
near the value $\lambda_1$. However, when $\lambda$ decreases, the
amplitude of oscillations becomes smaller and smaller, and this
critical line, or the boundary between phases 2 and 3, tends as a
whole to the point $(\mu_c,0)$ at $\lambda\to 0$, where $\mu_c$ is
the critical chemical potential at $L=\infty$. Moreover, it is clear
from Fig. 3 that at some values of $\lambda$ the SC phase is allowed
at even smaller values of $\mu$ (up to values $\mu\approx 0.2$ GeV)
than it occurs at $L=\infty$.

In Fig. 4 the behavior of the gap $\sigma$ vs $\lambda$ at
$\mu=0.18$ GeV is depicted (at this value of $\mu$ the gap $\Delta$
equals to zero). In Figs 5, 6 the gaps $\Delta$ and $\sigma$ vs
$\lambda$ are represented at $\mu=0.4$ GeV, correspondingly. As it
is clear from Figs 3--5, the oscillations both of the critical line
and gaps are characteristic features of the model in the finite
volume. Clearly, these quantities oscillate strongly vs $\lambda$.
One should also take into account that the gaps $\sigma$ and $\Delta$
from Figs 4-6 are really discontinuous functions vs $\lambda$ in the
points $\lambda_0$, $\lambda_1$, $\lambda_2$ etc.

Finally, let us discuss the influence of a nontrivial topology on
the values of particle density $n_\mu(\lambda)=-\partial\Omega_{L
a}/\partial\mu$ in the SC phase. In Fig. 7 its behavior vs $\lambda$
is presented at fixed $\mu =0.4$ GeV in comparison with the particle
density at $\lambda =0$ ($L=\infty$). It is clear from this figure
that at some finite values of $L$ which correspond to values of
$\lambda$ from rather small vicinities of $\lambda_k$ ($k=2,3,..$)
the superconducting phase is realized at smaller particle densities
than at $L=\infty$ (in these cases the ratio
$n_\mu(\lambda)/n_\mu(0)$ is less than 1) and at the same value of
$\mu =0.4$ GeV. It was mentioned above that the SC phase may occur
even at $\mu<\mu_c$ if $L$ is finite (see Fig. 3). It is interesting
to note that in these cases the particle density can also reach
rather small values. For example, it turns out that the point
$(\mu=0.21,\lambda=0.1255)$ GeV lies in the SC phase (see Fig. 3).
Numerical calculations show that for this set of parameters the
density of particles $n_\mu(\lambda)$ inside SC matter is much
smaller than $n_c$, where $n_c$ is the density at $\mu=\mu_c$ and
$L=\infty$, since in this case we have $n_\mu(\lambda)\approx
0.36~n_c$. 
These facts might be understood by taking into account the results of
the papers \cite{madsen,kiriyama}, where in particular it was shown
that the actual baryonic chemical potential, i.e. the energy per one
baryon, increases with decreasing size $L$ of a system. Hence, for
rather small values of the chemical potential $\mu$, for example at
$\mu<\mu_c$, by decreasing $L$, it is in principle possible, to reach
the value of the actual baryonic chemical potential at which the SC
gap $\Delta$ is opened, although in $R^3$-space the superconducting
phase is not realized at the same value of $\mu<\mu_c$. 

Note that the above mentioned as well as the following results should
be taken with caution when $\lambda \gtrsim\Lambda$. The reason is
that in this case the infrared cutoff $\lambda$ comes closer to the
ultraviolet one $\Lambda$, and thus the phase space is decreased due
to the presence of the $\theta$-function, e.g., in the expression
(\ref{18}).

\begin{figure}
 \includegraphics[width=0.45\textwidth]{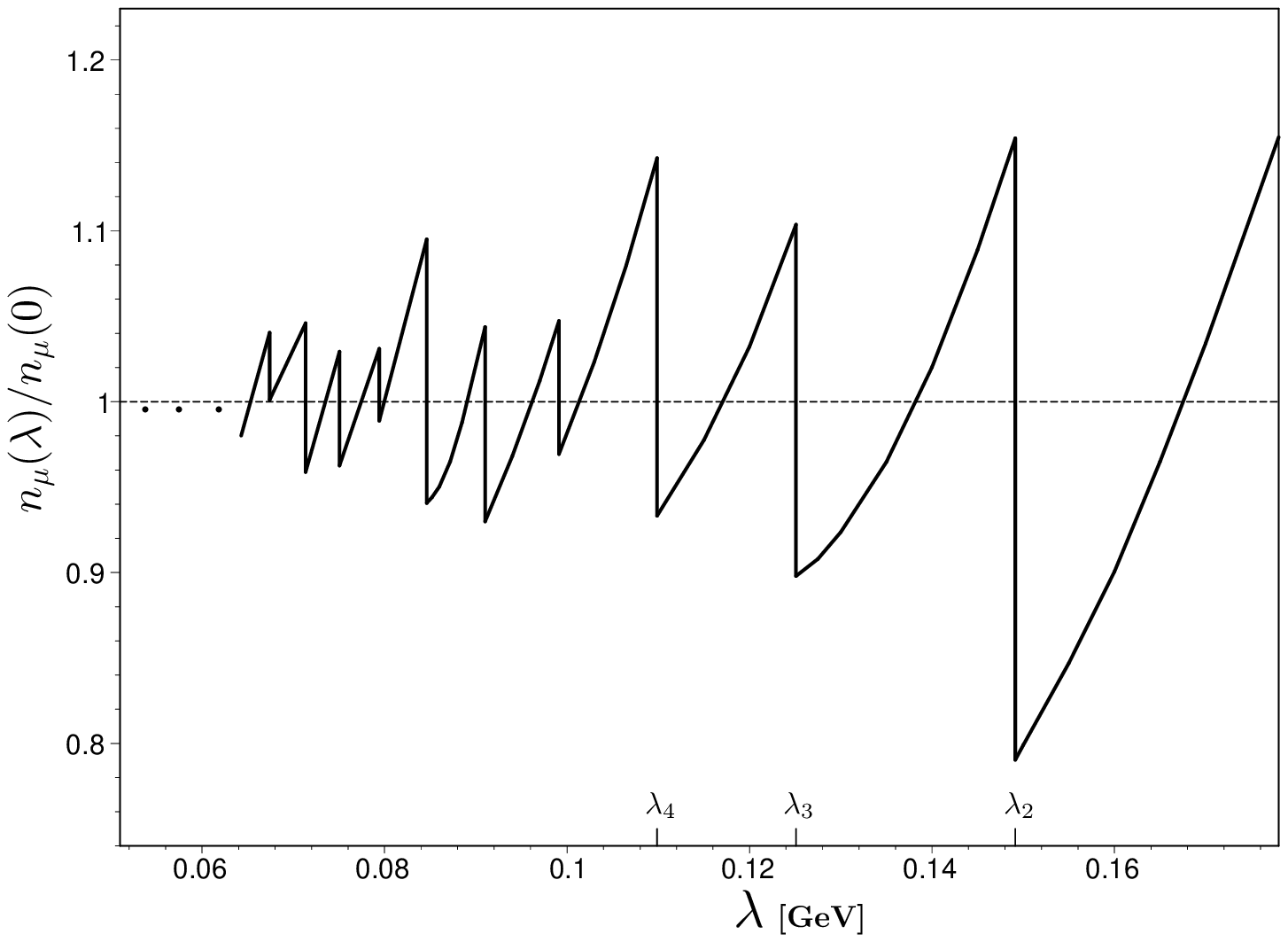}
 \hfill
 \includegraphics[width=0.45\textwidth]{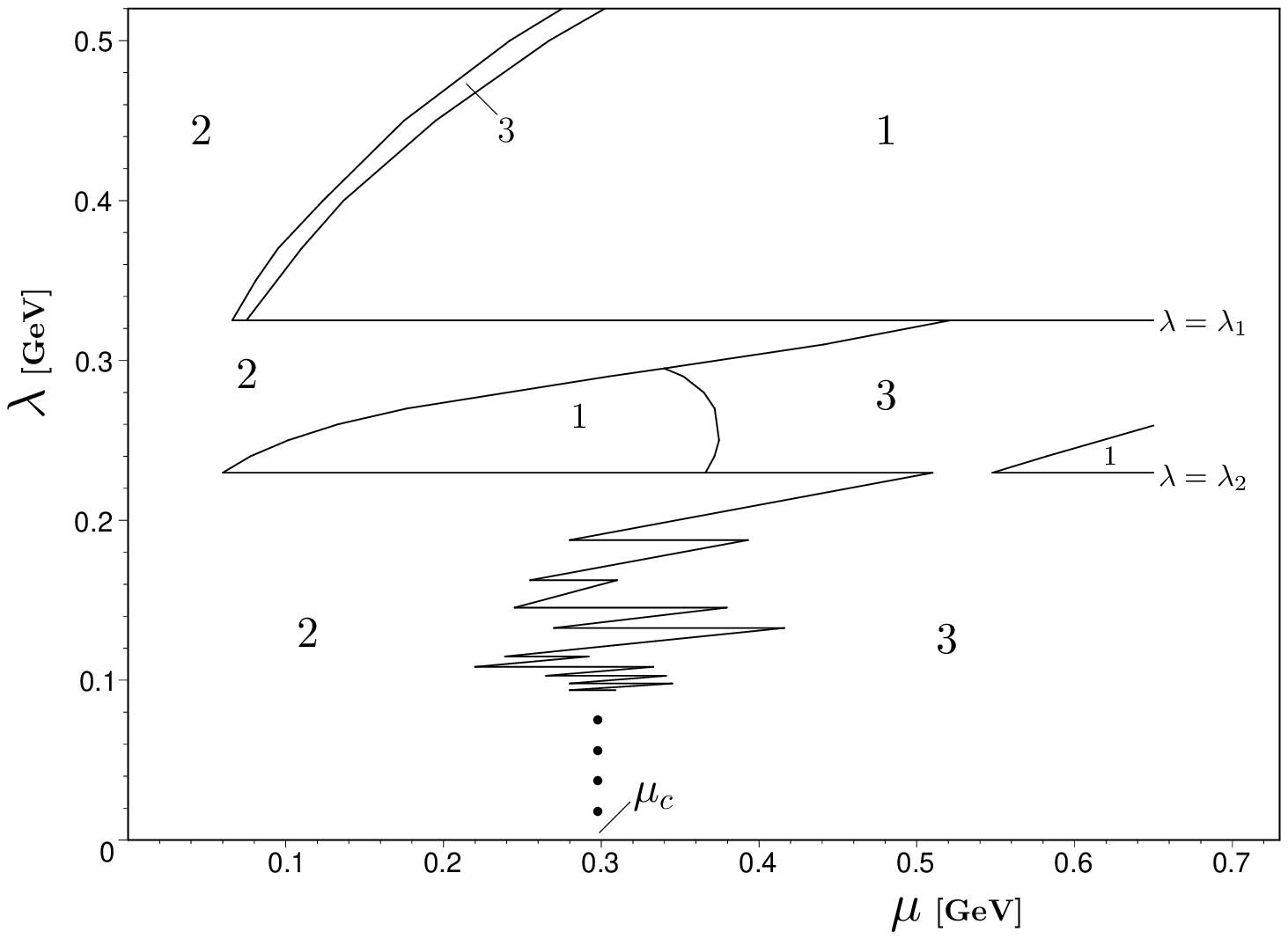}\\
\parbox[t]{0.45\textwidth}{ \caption{
Ratio of particle densities $\frac{n_\mu(\lambda)}{n_\mu(0)}$ vs
$\lambda$ at $\mu=0.4$ GeV  in the case of  $S^1\otimes S^1\otimes
S^1$ space topology with antiperiodic boundary conditions. (The
quantities $\lambda_k$ are presented in the text below
eq.(\ref{19}).) }
 }\hfill
 \parbox[t]{0.45\textwidth}{\caption{Phase structure in the case of
$S^1\otimes S^1\otimes S^1$ space topology with periodic boundary
conditions. Here $\lambda_1= 0.325$ GeV, $\lambda_2\approx 0.223$
GeV and other notations are given in Fig. 3.
 } }
\end{figure}
\label{anti}

\subsection{The case of periodic boundary conditions}

Applying the rule (\ref{17}) with $\alpha =0$ in the expression
(\ref{15}), we immediately obtain the TDP of the system in the
case of periodic (``p'') boundary conditions:
\begin{eqnarray}
\Omega_{L p}(\sigma,\Delta)&=&
\frac{\sigma^2}{4G}+\frac{\Delta^2}{4H}-
\frac{\lambda^3}{\pi^3}\sum_{i=-\infty}^\infty\sum_{k=-\infty}^\infty
\sum_{l=-\infty}^\infty
\Theta (\Lambda^2-p^2_{ip}-p^2_{kp}-p^2_{lp})\Big [{\cal E}_{\Delta
ikl}^{p+}+ {\cal E}_{\Delta ikl}^{p-}\Big ],\label{22}
\end{eqnarray}
where $\lambda=\pi/L$, ${\cal E}_{\Delta ikl}^{p\pm}=\sqrt{\left
({\cal E}_{ikl}\pm\mu\right )^2+\Delta^2}$, ${\cal
E}_{ikl}=\sqrt{p^2_{ip}+p^2_{kp}+p^2_{lp}+\sigma^2}$, and
$p_{ip}=2i\lambda$ etc. In the periodic case it is very convenient
to separate in each of the sums in (\ref{22}) the contributions from
zero modes, so the expression (\ref{22}) can be rearranged in the
following way:
\begin{eqnarray}
&&\Omega_{L p}(\sigma,\Delta)=
\frac{\sigma^2}{4G}+\frac{\Delta^2}{4H}-
\frac{\lambda^3}{\pi^3}\left [{\cal E}_{\Delta 000}^{p+}+ {\cal
E}_{\Delta 000}^{p-}\right ]-\frac{6\lambda^3}{\pi^3}
\sum_{i=1}^\infty \Theta (\Lambda^2-p^2_{ip})\Big [{\cal E}_{\Delta
i00}^{p+}+ {\cal E}_{\Delta i00}^{p-}\Big ]\nonumber\\
&-&\frac{12\lambda^3}{\pi^3} \sum_{i=1}^\infty\sum_{k=1}^\infty
\Theta (\Lambda^2-p^2_{ip}-p^2_{kp})\Big [{\cal E}_{\Delta
ik0}^{p+}+ {\cal E}_{\Delta ik0}^{p-}\Big ]
-\frac{8\lambda^3}{\pi^3}\sum_{i=1}^\infty\sum_{k=1}^\infty\sum_{l=1}
^\infty
\Theta (\Lambda^2-p^2_{ip}-p^2_{kp}-p^2_{lp})\Big [{\cal E}_{\Delta
ikl}^{p+}+ {\cal E}_{\Delta ikl}^{p-}\Big ].\label{23}
\end{eqnarray}
Now, as in the previous section, we interpret the quantities ${\cal
E}_{\Delta ikl}^{p-}$ (${\cal E}_{\Delta ikl}^{p+}$) in this
expression as the energies of the fermion (antifermion)
quasiparticles which can be labeled again by a triple of integers
$(i,k,l)$ with $i\ge k\ge l\ge 0$. Clearly, the quasiparticles of
the type $(0,0,0)$ always contribute to the expression (\ref{23}),
whereas the contribution of the other energy levels depends on
the $\lambda$-values. So, it is convenient, as in the case with
antiperiodic boundary conditions, to divide the $(\mu,\lambda)$
plane into strips
\begin{eqnarray}
\omega_1=\{(\mu,\lambda):~\lambda>\lambda_1\},~\omega_2=\{(\mu,
\lambda): ~\lambda_1>\lambda>\lambda_2\},~\cdots~,~
\omega_n=\{(\mu,\lambda):~\lambda_{n-1}>\lambda>\lambda_n\},~\dots,
\label{24}\end{eqnarray}
where for the real quantities $\lambda_k$ we use the same notations
as in Section \ref{anti}, which now however take other values, i.e.
$\lambda_1=\sqrt{\Lambda^2/4}=0.325$ GeV,
$\lambda_2=\sqrt{\Lambda^2/8}\approx 0.223$ GeV,
$\lambda_3=\sqrt{\Lambda^2/12}\approx 0.188$ GeV etc. \footnote{As
in the antiperiodic case, in the periodic one the real quantities
$\lambda_{ikl}=\sqrt{\Lambda^2/N_{ikl}}$, where
$N_{ikl}=4(i^2+k^2+l^2)$, might be associated with corresponding
energy level $(i,k,l)$. The real expressions
$\lambda_1>\lambda_2>\lambda_3>\cdots$ in (\ref{24}) are just the
quantities $\lambda_{ikl}$ arranged in a decreasing order.}
The regions $\omega_n$ in (\ref{24}) are constructed in such a way
that in the strip $\omega_1$ only the fermion and antifermion levels
$(0,0,0)$ contribute to the TDP (\ref{23}), where it looks like
\begin{eqnarray}
\Omega_{L p1}(\sigma,\Delta)&=&
\frac{\sigma^2}{4G}+\frac{\Delta^2}{4H}-\frac{\lambda^3}{\pi^3}
\left [\sqrt{(\sigma +\mu)^2+\Delta^2}+
\sqrt{(\sigma-\mu)^2+\Delta^2}\right ].\label{25}
\end{eqnarray}
(In this region there are no terms of the infinite sums of
(\ref{23}) which supply a nonzero contributions to the TDP.) In the
strip $\omega_2$ the energy levels $(1,0,0)$  of fermion and
antifermion quasiparticles are switched on in addition, in the strip
$\omega_3$ the energy levels $(1,1,0)$ are switched on in addition,
etc. Investigating the global minimum point behavior of the TDP
(\ref{23}) in each of the strips (\ref{24}), it is possible to
obtain the phase structure of the model in the whole $(\mu,\lambda)$
plane, which is presented in Fig. 8. The behavior of the gaps
$\sigma$, $\Delta$ and particle density $n_\mu(\lambda)$ vs
$\lambda$ in the periodic case are qualitatively the same as in the
case with antiperiodic boundary conditions (see Figs 4-7).
\begin{figure}
 \includegraphics[width=0.45\textwidth]{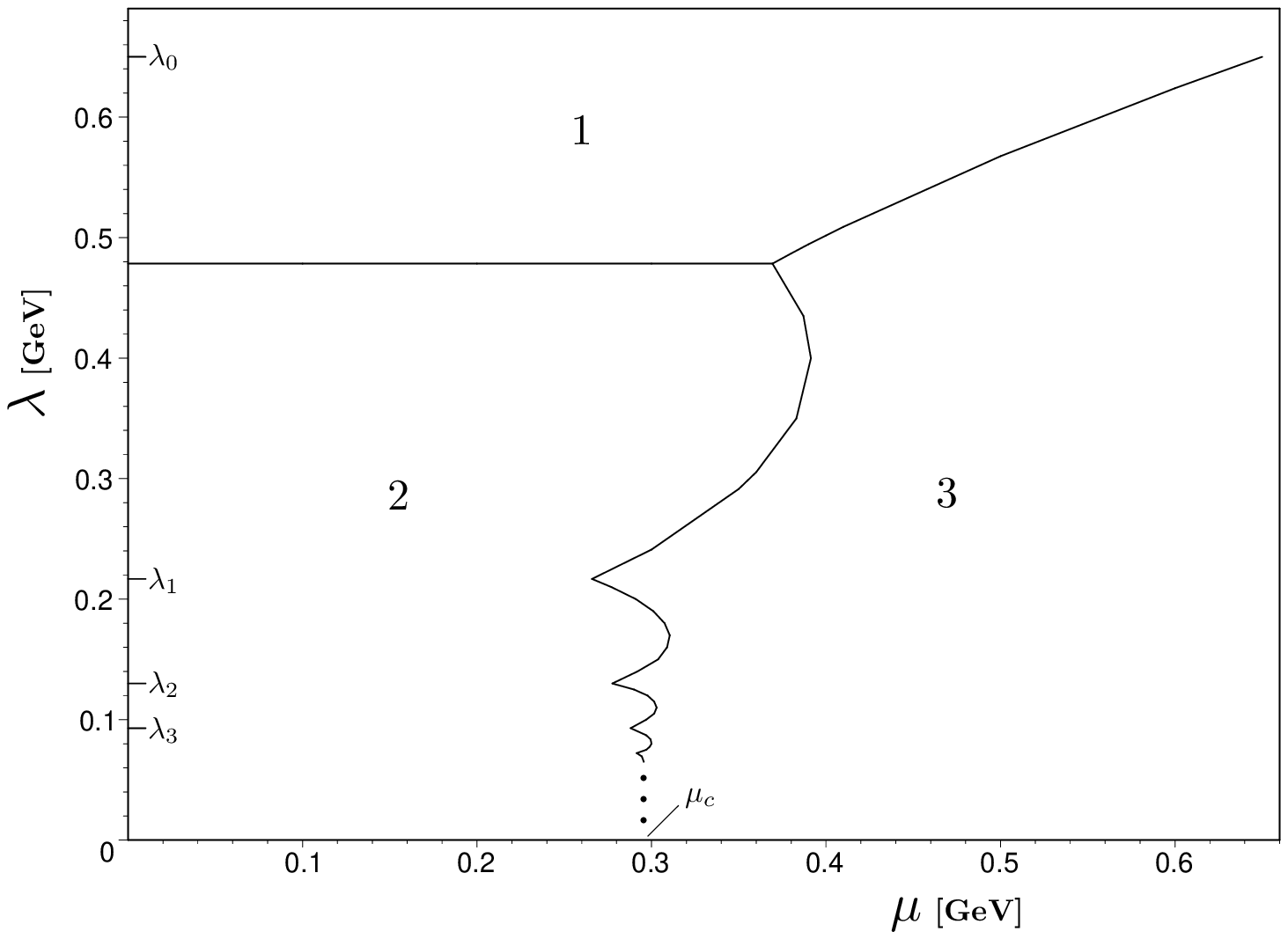}
 \hfill
 \includegraphics[width=0.45\textwidth]{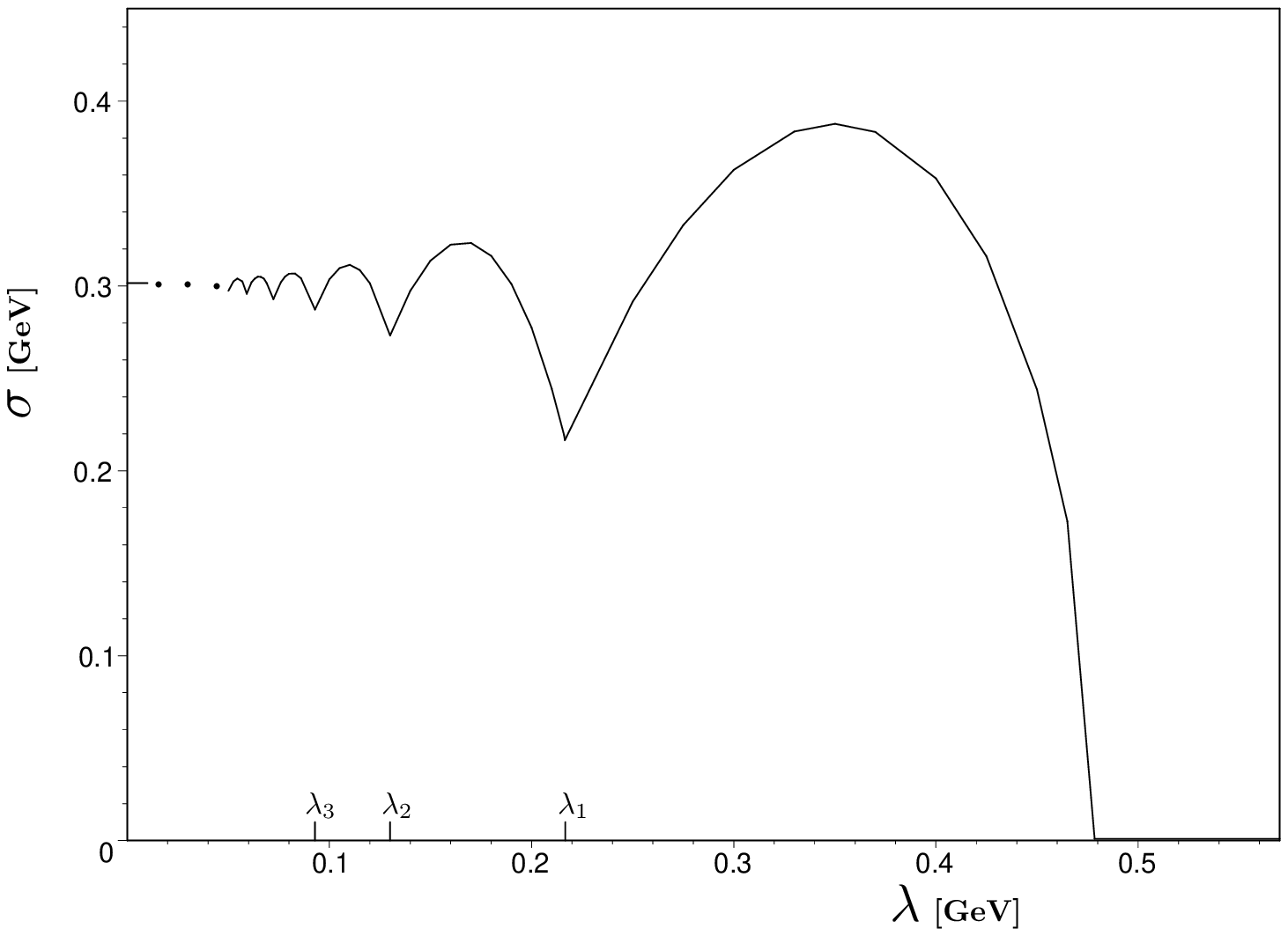}\\
\parbox[t]{0.45\textwidth}{ \caption{Phase structure in the case of
$R^2\otimes S^1$ space topology with antiperiodic boundary
conditions. Here $\lambda_k=\Lambda/(2k+1)$ ($k=0,1,2,3$) and other
notations are given in Fig. 3. } }\hfill
 \parbox[t]{0.45\textwidth}{\caption{The gap $\sigma$
vs $\lambda$ at $\mu=0.2$ GeV  in the case of $R^2\otimes S^1$ space
topology with antiperiodic boundary conditions. Here
$\lambda_k=\Lambda/(2k+1)$ ($k=1,2,3$).} }
\end{figure}
\label{period}

\section{The case of  $R^2\otimes S^1$ space topology}

In the present section we continue the investigation of the
difermion condensation in the spaces with nontrivial topology, this
time when it is of the form $R^2\otimes S^1$. For simplicity, it is
supposed here that the $z$-axis is compactified and fermion fields
satisfy some boundary conditions of the form (the $x,y$ coordinates
are not restricted):
\begin{eqnarray}
\psi_k(t,x,y,z+L)=e^{i\pi\alpha}\psi_k(t,x,y,z). \label{26}
\end{eqnarray}
As in the previous section, we shall use only two values of the
parameter $\alpha$: $\alpha=0$ for the periodic boundary condition
and $\alpha=1$ for the antiperiodic one. Recall that $L$ is the
length of the circumference $S^1$. Note also that the consideration
of any physical system in the above mentioned space topology is
equivalent to a restriction of the system inside an infinite layer
with thickness $L$. In this case, to obtain the thermodynamic
potential $\Omega_{L\alpha}(M,\Delta)$ of the initial system, one
again simply replaces the integration over $p_3$ in (\ref{15}) by an
infinite series, using the analogous rule:
\begin{eqnarray}
\int_{-\infty}^{\infty}\frac{dp_3}{2\pi}f(p_{3})\to\frac
1L\sum_{n=-\infty}^{\infty}f(p_{n\alpha}),~~~~p_{n\alpha}=
\frac{\pi}{L}(2n+\alpha),~~~n=0,\pm 1, \pm 2,... \label{27}
\end{eqnarray}

\subsection{The case of antiperiodic boundary conditions}

Applying the rule (\ref{27}) with $\alpha =1$ in (\ref{15}), one has
for the TDP $\Omega_{L a}$ of dense cold matter in a space of
$R^2\times S^1$ topology with antiperiodic boundary conditions the
following expression
\begin{eqnarray}
\Omega_{L a}(\sigma,\Delta)&=&
\frac{\sigma^2}{4G}+\frac{\Delta^2}{4H}-
\frac{2\lambda}{\pi}\sum_{i=0}^\infty\int\frac{d^2p}{(2\pi)^2}
\Theta (\Lambda^2-\vec p^2-p^2_{ia})\Big [{\cal E}_{\Delta i}^{a+}+
{\cal E}_{\Delta i}^{a-}\Big ],\label{28}
\end{eqnarray}
where $\lambda=\pi/L$, ${\cal E}_{\Delta i}^{a\pm}=\sqrt{\left
({\cal E}_{i}\pm\mu\right )^2+\Delta^2}$, ${\cal E}_{i}=\sqrt{\vec
p^2+p^2_{ia}+\sigma^2}$, and $p_{ia}=\lambda (2i+1)$. Recall that
the quantities ${\cal E}_{\Delta i}^{a-}$ (${\cal E}_{\Delta
i}^{a+}$) are the energies of fermion (antifermion) quasiparticles,
which are labeled by a discrete index $i$ ($i=0,1,2...$) and by a
continuous quantity $|\vec p|$. Integrating in (\ref{28}) over
two-dimensional momenta $\vec p$, we obtain
\begin{eqnarray}
\Omega_{La}(\sigma,\Delta) =\frac{\sigma^2}{4G}+\frac{\Delta^2}{4H}
+\frac{\lambda}{6\pi^2}\Theta\left (\frac{\Lambda
-\lambda}{2\lambda}\right )\Phi_{a}(\sigma,\Delta), \label{29}
\end{eqnarray}
where
\begin{eqnarray}
\Phi_{a}(\sigma,\Delta)&=&(N_a+1)\left\{3\mu \left
(\sqrt{\Lambda^2+\sigma^2}+\mu\right )\sqrt{\left
(\sqrt{\Lambda^2+\sigma^2}+\mu\right )^2+\Delta^2}-2\left [\left
(\sqrt{\Lambda^2+\sigma^2}+\mu\right )^2+\Delta^2\right
]^{3/2}\right\}\nonumber\\
&+&\sum_{n=0}^{N_a}\left\{2 \left [\left
(\sqrt{p^2_{na}+\sigma^2}+\mu\right )^2 +\Delta^2\right ]^{3/2}-3\mu
\left (\sqrt{p^2_{na}+\sigma^2}+\mu\right )\sqrt{\left
(\sqrt{p^2_{na}+\sigma^2}+\mu\right
)^2+\Delta^2}\right.\nonumber\\
&+&\left.3\mu\Delta^2\ln\left |\frac{\sqrt{\Lambda^2+\sigma^2}+\mu+
\sqrt{\left (\sqrt{\Lambda^2+\sigma^2}+\mu\right
)^2+\Delta^2}}{\sqrt{p^2_{na}+\sigma^2}+\mu+\sqrt{ \left
(\sqrt{p^2_{na}+\sigma^2}+\mu\right )^2+\Delta^2}}\right
|\right\}+(\mu\to -\mu), \label{30}
\end{eqnarray}
and $N_a\equiv \left [\frac{\Lambda -\lambda}{2\lambda}\right ]$
(recall, $[x]$ means the integer part of a real number $x$).

It is clear from the structure of the TDP (\ref{28})--(\ref{30})
that to obtain the behavior of its global minimum point vs $\mu$ and
$\lambda$ it is very convenient to divide again the plane
$(\mu,\lambda)$ into an infinite set of strips (\ref{20}), this time
with $\lambda_k=\Lambda/(2k+1)$. \footnote{One should not become
confused by the fact that we use the same notation for the
boundaries $\lambda_k$ of these strips in different cases. Indeed,
as it is clear from the text, the value of each $\lambda_k$ depends
strongly on both the space topology and boundary conditions.} So,
due to the presence of the $\Theta$-function in (\ref{29}), in the
region $\omega_0$, i.e. at $\lambda>\lambda_0\equiv\Lambda$, the TDP
(\ref{28})-(\ref{29}) has a trivial form, i.e.
\begin{eqnarray}
\Omega_{L a}(\sigma,\Delta)\Big |_{(\mu,\lambda)\in\omega_0}=
\frac{\sigma^2}{4G}+\frac{\Delta^2}{4H},\label{31}
\end{eqnarray}
whose global minimum point $(\sigma =0,\Delta =0)$ corresponds to
the symmetric phase. In the region $\omega_1$, i.e. at
$\lambda_0>\lambda>\lambda_1$, the integer $N_a$ from (\ref{30}) is
equal to zero, hence only  quasiparticles with the energies ${\cal
E}_{\Delta 0}^{a\pm}$ contribute to the TDP in this case. Studying
the behavior of the GMP of the TDP vs $\mu$ and $\lambda$, we
conclude that in the strip $\omega_1$ three phases, the symmetric
phase 1, the phase with broken chiral symmetry 2, and the
superconducting phase 3, occur (see the part of Fig. 9 that
corresponds to $\lambda_0>\lambda>\lambda_1$). In the strip
$\omega_2$ we have in the expression (\ref{30}) $N_a=1$, so here the
quasiparticle energies ${\cal E}_{\Delta 1}^{a\pm}$ contribute to
the TDP in addition etc. However, our numerical calculations show
that in each of the strips $\omega_2$, $\omega_3$,...  only two
phases, the phase 2 and the SC phase 3, of the model are realized.
The boundary between the phase with broken chiral symmetry 2 and the
superconducting one 3 again oscillates when $\lambda\to 0$ (see Fig.
9). It turns out that in addition to the critical curve the gaps
$\sigma$ and $\Delta$ oscillate vs $\lambda$ at fixed values of
$\mu$ (see Figs 10, 11). Comparing these figures with Figs 4, 5, we
see that the more dimensions are compactified, the more sharp
oscillations of physical quantities occur. In particular, it is
clear from Figs 10, 11 that in the case under consideration the gaps
$\sigma$ and $\Delta$ are some continuous functions vs $\lambda$,
whereas in the case with $S^1\otimes S^1\otimes S^1$ space topology
these quantities have discontinuities in the points $\lambda_0$,
$\lambda_1$, etc.

Finally, in Fig. 12 the ratio of particle densities
$\frac{n_\mu(\lambda)}{n_\mu(0)}$ vs $\lambda$ inside the SC phase
is presented at $\mu=0.4$ GeV. It is clear that in small vicinities
around $\lambda_k$ ($k=2,3,..$) the density at $\lambda\ne 0$ is
less than the particle density at $\lambda=0$. It seems intuitively
clear that the smaller a particle density is, the
easier a corresponding state of the system might be created. So to
reduce the efforts in obtaining the SC phase, one could simply fix
the $\lambda$-parameter not far from one of the $\lambda_k$-values.
Hence finite size effects might promote the transition of a physical
system into its superconducting phase.
\begin{figure}
 \includegraphics[width=0.45\textwidth]{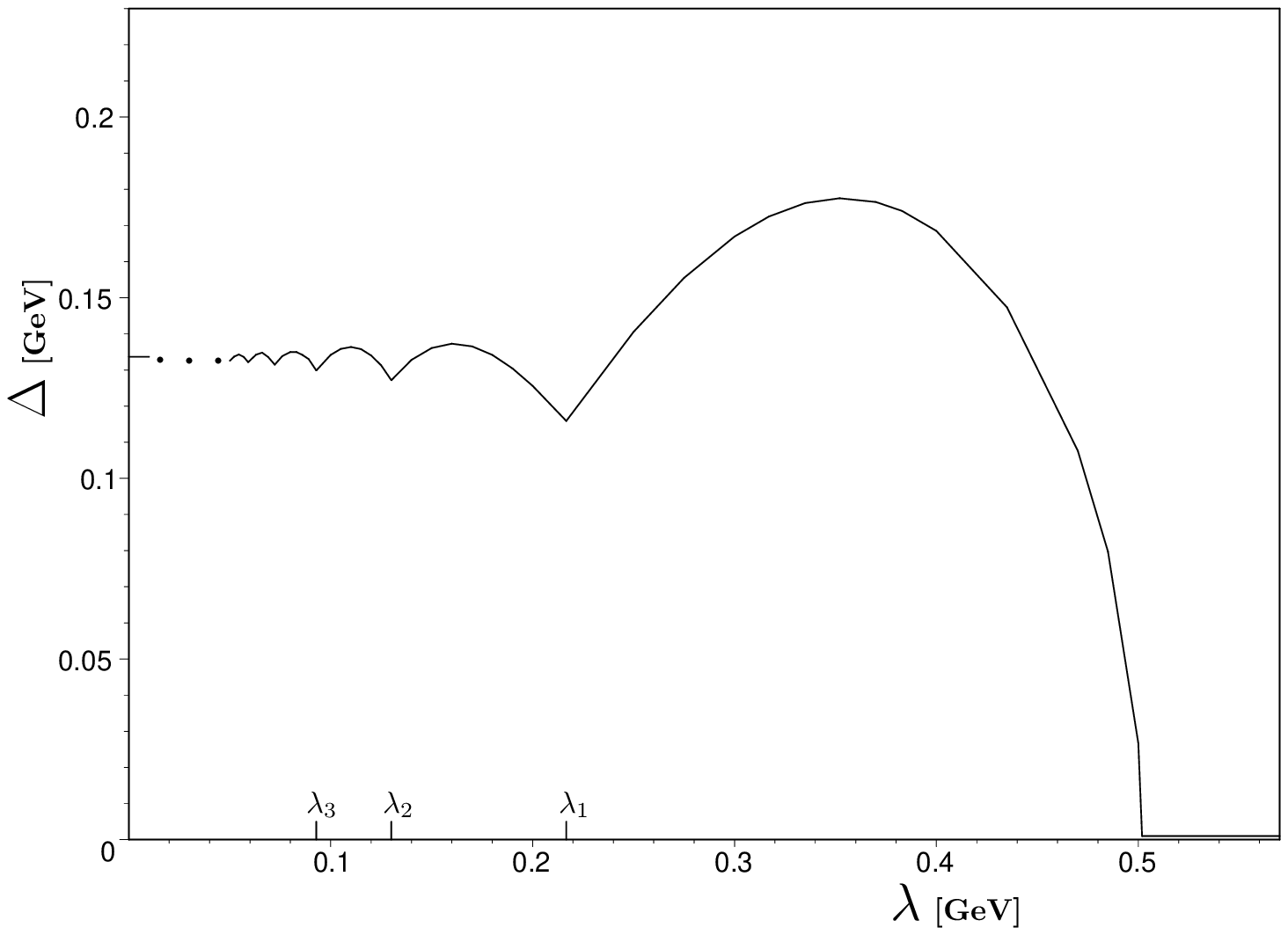}
 \hfill
 \includegraphics[width=0.45\textwidth]{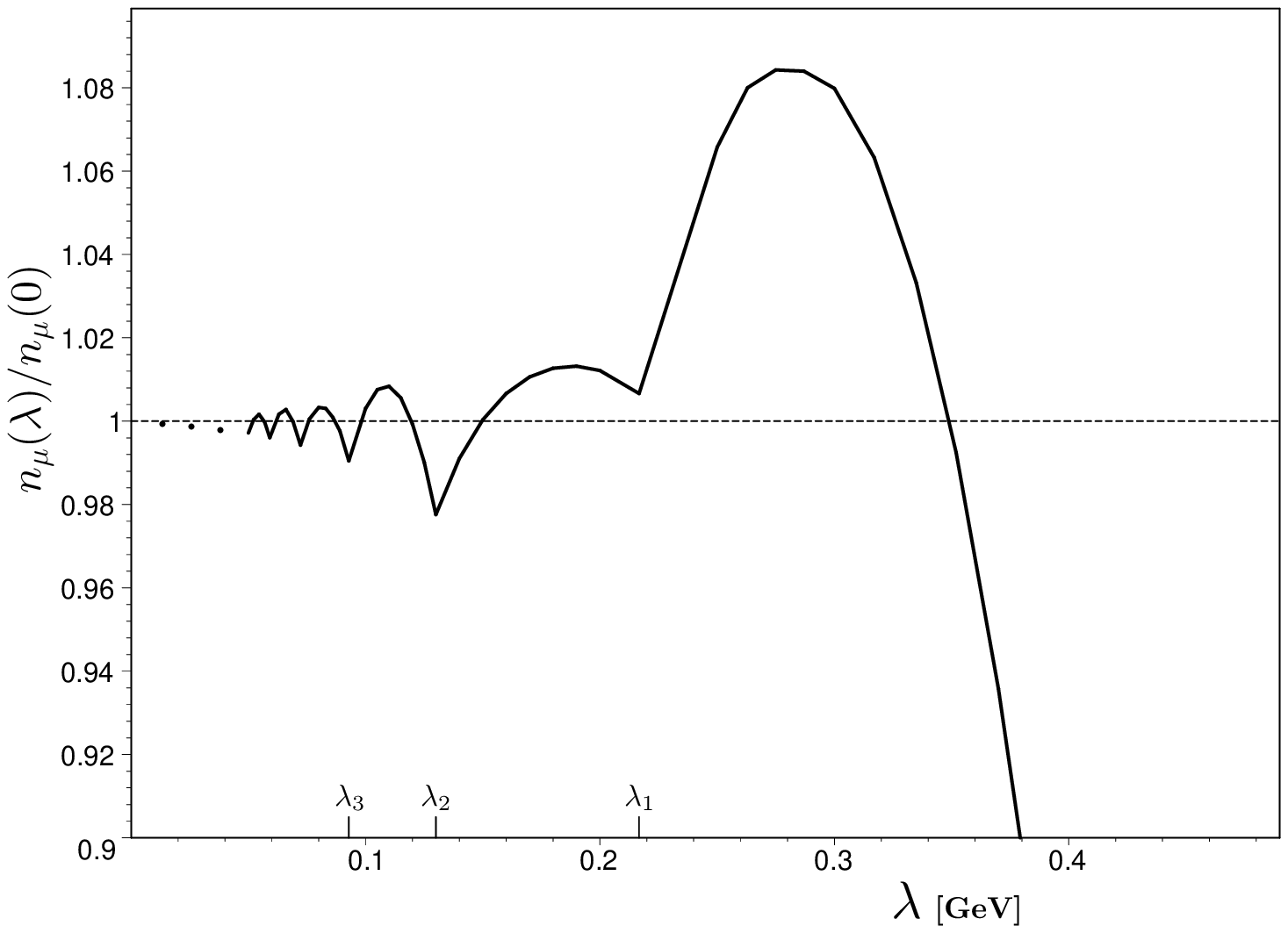}\\
\parbox[t]{0.45\textwidth}{ \caption{The gap $\Delta$
vs $\lambda$ at $\mu=0.4$ GeV  in the case of $R^2\otimes S^1$ space
topology with antiperiodic boundary conditions.  Here
$\lambda_k=\Lambda/(2k+1)$ ($k=1,2,3$).}}\hfill
 \parbox[t]{0.45\textwidth}{\caption{Ratio of particle densities
 $\frac{n_\mu(\lambda)}{n_\mu(0)}$ vs $\lambda$ at $\mu=0.4$ GeV in
 the case of  $R^2\otimes S^1$ space topology with antiperiodic
 boundary  conditions.  Here $\lambda_k=\Lambda/(2k+1)$ ($k=1,2,3$).}
 }
\end{figure}

\subsection{The case of periodic boundary conditions}

Clearly, to obtain the TDP of the system $\Omega_{L p}$ in this
case, it is necessary to use in (\ref{15}) the rule (\ref{27}) with
$\alpha =0$. As a result, we have
\begin{eqnarray}
\hspace{-7mm}\Omega_{Lp}(\sigma,\Delta)
=\frac{\sigma^2}{4G}+\frac{\Delta^2}{4H}&-&\frac{\lambda}{\pi}
\int\frac{d^2p}{(2\pi)^2} \theta (\Lambda^2-|\vec p|^2)\Big [{\cal
E}_{\Delta 0}^{p+}+{\cal
E}_{\Delta 0}^{p-}\Big ]\nonumber\\
&-&\frac{2\lambda}{\pi}\sum_{n=1}^\infty\int\frac{d^2p}{(2\pi)^2}
\theta\left (\Lambda^2-|\vec p|^2-p^2_{np}\right )\Big [{\cal
E}_{\Delta n}^{p-}+{\cal E}_{\Delta n}^{p+}\Big ], \label{2000}
\end{eqnarray}
where $\lambda=\pi/L$, ${\cal E}_{\Delta n}^{p\pm}=\sqrt{\left
({\cal E}_{n}\pm\mu\right )^2+\Delta^2}$, ${\cal E}_{n}=\sqrt{\vec
p^2+p^2_{np}+\sigma^2}$, $p_{np}=2n\lambda$ ($n=0,1,2...$). Recall
that the quantities ${\cal E}_{\Delta n}^{p-}$ (${\cal E}_{\Delta
n}^{p+}$) are the energies of fermion (antifermion) quasiparticles,
which are labeled by a discrete index $n$ ($n=0,1,2...$) and by a
continuous quantity $|\vec p|$, in addition. We find it convenient
to separate in (\ref{2000}) the contribution from the zero modes,
i.e. the contribution from quasiparticles with $n=0$. Integrating in
(\ref{2000}) over two-dimensional momenta $\vec p$, we have
\begin{eqnarray}
&&\hspace{-7mm}\Omega_{Lp}(\sigma,\Delta)
=\frac{\sigma^2}{4G}+\frac{\Delta^2}{4H}+\frac{\lambda}{12\pi^2}
F(\sigma,\Delta)+ \frac{\lambda}{6\pi^2}\Theta\left
(\frac{\Lambda}{2\lambda}-1\right )\Phi_p(\sigma,\Delta),
\label{2002}
\end{eqnarray}
where
\begin{eqnarray}
\hspace{-7mm}F(\sigma,\Delta)&=&2\left
[(\sigma+\mu)^2+\Delta^2\right ]^{3/2}-2\left [\left
(\sqrt{\Lambda^2+\sigma^2}+\mu\right )^2+\Delta^2\right
]^{3/2}\nonumber\\
&+&3\mu \left (\sqrt{\Lambda^2+\sigma^2}+\mu\right ) \sqrt{\left
(\sqrt{\Lambda^2+\sigma^2}+\mu\right )^2+\Delta^2}-3\mu (\sigma+\mu
)\sqrt{(\sigma+\mu)^2+\Delta^2}\nonumber\\
&+&3\mu\Delta^2\ln\left |\frac{\sqrt{\Lambda^2+\sigma^2}+\mu+
\sqrt{\left (\sqrt{\Lambda^2+\sigma^2}+\mu\right
)^2+\Delta^2}}{\sigma+\mu+\sqrt{(\sigma+\mu)^2+\Delta^2}}\right
|+(\mu\to -\mu),\label{2001}\\
\Phi_p(\sigma,\Delta)&=&N_p\left\{3\mu \left
(\sqrt{\Lambda^2+\sigma^2}+\mu\right )\sqrt{\left
(\sqrt{\Lambda^2+\sigma^2}+\mu\right )^2+\Delta^2}-2\left [\left
(\sqrt{\Lambda^2+\sigma^2}+\mu\right )^2+\Delta^2\right
]^{3/2}\right\}\nonumber\\
&+&\sum_{n=1}^{N_p}\left\{2 \left [\left
(\sqrt{p^2_{np}+\sigma^2}+\mu\right )^2 +\Delta^2\right ]^{3/2}-3\mu
\left (\sqrt{p^2_{np}+\sigma^2}+\mu\right )\sqrt{\left
(\sqrt{p^2_{np}+\sigma^2}+\mu\right
)^2+\Delta^2}\right.\nonumber\\
&+&\left.3\mu\Delta^2\ln\left |\frac{\sqrt{\Lambda^2+\sigma^2}+\mu+
\sqrt{\left (\sqrt{\Lambda^2+\sigma^2}+\mu\right
)^2+\Delta^2}}{\sqrt{p^2_{np}+\sigma^2}+\mu+\sqrt{ \left
(\sqrt{p^2_{np}+\sigma^2}+\mu\right )^2+\Delta^2}}\right
|\right\}+(\mu\to -\mu), \label{2003}
\end{eqnarray}
and $N_p\equiv\left [\frac{\Lambda}{2\lambda}\right ]$ is the
integer part of the real number in the square bracket.
\begin{figure}
\includegraphics[width=0.45\textwidth]{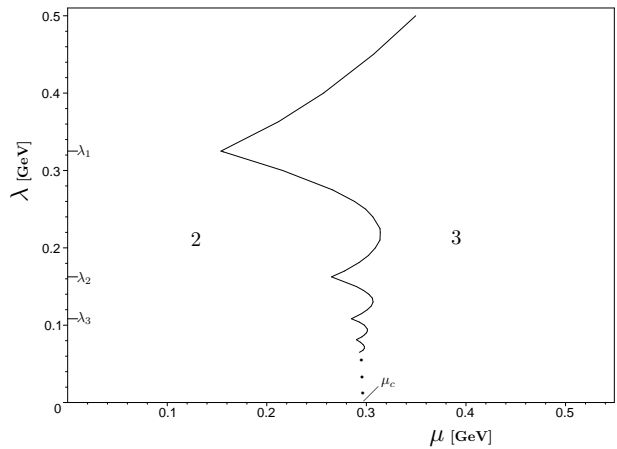}
 \caption{Phase structure in the case of $R^2\otimes S^1$ space
 topology with periodic boundary conditions. Here
 $\lambda_k=\Lambda/(2k)$ ($k=1,2,3$),
and the numbers 2,3 denote the chirally broken and SC phases. }
 \end{figure}

As in section \ref{period}, we again divide the
$(\mu,\lambda)$-plane into an infinite set of strips $\omega_k$
(\ref{24}), where in the present case we have
$\lambda_k=\Lambda/(2k)$ ($k=1,2,...$). Then, in the region
$\omega_1$ the zero energy levels ${\cal E}_{\Delta 0}^{p\pm}$ of
quasiparticles contribute to the TDP (\ref{2000}), where it looks
like
\begin{eqnarray}
\Omega_{L p}(\sigma,\Delta)\Big |_{(\mu,\lambda)\in\omega_1}=
\frac{\sigma^2}{4G}+\frac{\Delta^2}{4H}+\frac{\lambda}{12\pi^2}
F(\sigma,\Delta)\label{2004}
\end{eqnarray}
(here $F(\sigma,\Delta)$ is given in (\ref{2001})). In the region
$\omega_2$ the next energy levels ${\cal E}_{\Delta 1}^{p\pm}$ are
switched on in addition, so in the expression for the function
$\Phi_p(\sigma,\Delta)$ (\ref{2003}) we have $N_p=1$, etc. It turns
out that with each region $\omega_k$ ($k\ge 2$) the value $N_p=k-1$
in (\ref{2003}) is associated. As a result, in $\omega_k$ all the
quasiparticle energy levels with quantum numbers $n=0,1,..,, k-1$
contribute to the expression for the TDP. Studying numerically
step-by-step the behavior of the GMP of the TDP (\ref{2002}) in the
strips $\omega_0$, $\omega_1$, $\omega_2$,..., we obtain the
$(\mu,\lambda)$ phase portrait of the initial NJL model which is
presented in Fig. 13. In contrast to the phase portrait in the
antiperiodic case (see Fig. 9), it has only two phases, the chirally
broken phase 2 and the superconducting one 3.

Notice that in the case with periodic boundary conditions the
behavior of the gaps $\sigma$, $\Delta$ and particle density
$n_\mu(\lambda)$ are qualitatively the same as in the antiperiodic
case (see Figs 10 -- 12).

\section{Summary and discussion}

In the present paper we have investigated the influence of
finite-size effects on the superconductivity (SC) phenomenon which
might exist in dense cold fermionic matter. Note that in the SC
phase the $U(1)$ charge group is spontaneously broken down due to
Cooper pair (difermion) condensation. It is well-known that quantum
fluctuations of fields can destroy symmetry breaking in a finite
volume, so studying this effect requires to be sure that such
fluctuations and their corresponding next to leading order
corrections will not spoil the leading order mean field results.
Concerning the application of usual QCD-like NJL models to the
description of CSC in the mean field approximation, a small
perturbative expansion parameter guaranteeing the suppression of
quantum fluctuations is absent. Thus, in this case there is no
confidence that quantum fluctuations are in general negligible and,
particularly in the case of finite systems, that they could not
destroy color symmetry breaking. These obstacles were the main
reason why we decided to consider instead of a QCD-like quark model
the ``toy'' NJL model (1) for fermions which seems to us technically
more adequate for studying the influence of finite volume effects on
the Cooper pairing. In fact, in this model at large $N$ a small
expansion parameter $1/N$ appears, so the next to leading order
corrections in $1/N$ are certainly negligible. By this reason,
quantum fluctuations of fields become suppressed and cannot destroy
spontaneous symmetry breaking for the considered fermion system in a
finite volume (see also the discussion in Introduction).

Let us summarize in more detail some of the main results. First, it
was shown in the leading order of the $1/N$-expansion that at
$T=0,L=\infty$ ($L$ is the linear size of the system) there is a
phase transition in the considered fermion model from the chiral
symmetry breaking phase 2 to a superconducting one 3 at the critical
value of the chemical potential $\mu=\mu_c\approx 0.3$ GeV. Next, we
have studied in the leading order over $1/N$ the phase structure of
this model both in the space with topology $S^1\otimes S^1\otimes
S^1$ and $R^2\otimes S^1$ taking into account periodic and
untiperiodic boundary conditions for fermion fields. It turns out
that for all cases the critical line between phase 2 and 3 as well as
the gaps $\sigma$, $\Delta$ and the particle density $n_\mu(\lambda)$
are oscillating functions vs $\lambda\sim 1/L$. Generally we found
that the more spatial dimensions are compactified, the stronger
oscillations occur.

Secondly, it is interesting to note that at
finite $L$ the superconducting phase 3 might be realized at
sufficiently smaller values of the chemical potential $\mu$ and
particle density $n_\mu(\lambda)$, than at
$L=\infty$. Indeed, as it is clear from, e.g., Figs. 3, 13 for some
values of $\lambda$ the phase 3 occurs at $\mu=0.2$ GeV or even
smaller values. Moreover, as it is clear from the discussion at the
end of section \ref{anti}, for such sufficiently small values of
$\mu$ the particle density inside the SC phase might be as small as
$\approx 0.4~n_c$, where $n_c$ is the density at which SC is realized
at $L=\infty$. Hence, the compactification procedure itself might
shift the SC phase transition to smaller particle densities.

Hopefully, the above investigations could motivate other studies
of finite size effects and eventually find some physical
applications.

\section*{Acknowledgments}

One of us (K.G.K.) is grateful to Prof. M. Mueller-Preussker and his
colleagues for the kind hospitality at the Institute of Physics of
the Humboldt-University during November -- December of 2009 and to
{\it Deutscher Akademischer Austauschdienst} (DAAD) for financial
support.

\end{document}